
 \input harvmac
\input epsf.tex
\def\caption#1{{\it
	\centerline{\vbox{\baselineskip=12pt
	\vskip.15in\hsize=4.2in\noindent{#1}\vskip.1in }}}}
\def\pyidk{PHY-9057135}

   \def\CL{{\cal L}}
  \def\CO{{\cal O}}

%
%
%
%
%

\def\bar#1{\overline{#1}}

\def\bra#1{\left\langle #1\right|}
\def\ket#1{\left| #1\right\rangle}
\def\braket#1#2{\bra{#1}#2\rangle}

\def\half{{\textstyle{1\over2}}} 
\def\frac#1#2{{\textstyle{#1\over #2}}} 
%
%
%
%

\def\Tr{\mathop{\rm Tr}}

%
%
%
%

%
%
\def\ltap{\ \raise.3ex\hbox{$<$\kern-.75em\lower1ex\hbox{$\sim$}}\ }
\def\gtap{\ \raise.3ex\hbox{$>$\kern-.75em\lower1ex\hbox{$\sim$}}\ }
\def\gl{\ \raise.5ex\hbox{$>$}\kern-.8em\lower.5ex\hbox{$<$}\ }
\def\roughly#1{\raise.3ex\hbox{$#1$\kern-.75em\lower1ex\hbox{$\sim$}}}
%
%

\def\etal{\hbox{\it et al.\ }}

\def\np#1#2#3{{Nucl. Phys. } B{#1} (#2) #3}
\def\npa#1#2#3{{Nucl. Phys. } A{#1} (#2) #3}
\def\pl#1#2#3{{Phys. Lett. } {#1}B (#2) #3}
\def\prl#1#2#3{{Phys. Rev. Lett. } {#1} (#2) #3}
\def\physrev#1#2#3{{Phys. Rev. } {#1} (#2) #3}

\relax
\def\Dsl{\,\raise.15ex \hbox{/}\mkern-13.5mu D}
\def\vsl{\,\raise.15ex \hbox{/}\mkern-10.5mu v}
\def\frac#1#2{{\textstyle{#1 \over #2}}}

\def\[{\left[}
\def\]{\right]}
\def\({\left(}
\def\){\right)}
\def\lfb{\bigskip\noindent}
\def\lfm{\medskip\noindent}
\def\lfs{\smallskip\noindent}
\noblackbox

\Title{\vbox{
\hfill DOE/ER/40561-205-INT95-00-92 \medskip \hfill UW/PT 95-05}}
{\vbox{\centerline{Effective Field Theories}}}
\medskip
\centerline{{\sl Lectures given at the Seventh Summer School in Nuclear
Physics:\  ``Symmetries''} }
\centerline{{\sl Institute for Nuclear Theory, June 19 - 30, 1995}}
\bigskip
\vskip .5in
\centerline{David B. Kaplan\footnote{$^a$}{Email:
{\tt dbkaplan@phys.washington.edu}}}
 \centerline{{\sl
Institute for
Nuclear Theory, NK-12}}
\centerline{{\sl University of Washington}}
\centerline{{\sl  Seattle WA 98195, USA}}
\vfill
{ Three lectures in which I give an introduction to effective field
theories  in particle physics. }
\Date{6/95}
\baselineskip 18pt

\newsec{Introduction}
Physics, chemistry and biology are all sciences describing different aspects of
the same world, yet their practice differs enormously...why is that?  To a
large extent the cultural differences between the fields can be explained in
terms of fundamental difference between the physical systems being analyzed:
A  biological system typically has  many unrelated energy scales         of
comparable magnitude, such as the low lying excitation spectra of various
different enzymes collaborating in a biological process.  While it is possible
to crudely explain why the body temperature of a human is $\CO(10^2)\ K$ in
terms the fine structure constant $\alpha$ and the masses of the proton and the
electron,  you will not be able to explain why a human is healthy with a
temperature of $310.5\ K$ but not with a temperature of $320.5\ K$ without a
thorough understanding of human physiology.

In contrast, a  physics problem  tends to involve  energy scales that are
widely separated, which allows one --- with care --- to determine many of the
properties of a system using the tool of dimensional analysis.  To see how this
works, we first choose physical units with the speed of light and Planck's
constant set to unity:
$$\hbar = c = 1\ .$$
Then all physical parameters can be said to have the dimension of mass to some
power.  In particular, if some quantity $X$ has dimensions $[mass]^n$, we will
just say ``$X$ has dimension $n$'' or $[X]=n$.  You should be able to convince
yourself that
$$[volume] \sim [\int {d^3 x} ]=-3$$
$$[G_N] \sim [G_F] =-2$$
$$[length] \sim [time] \sim [\mu_N]=-1$$
$$[velocity] \sim [\alpha]=0$$
$$[energy] \sim[momentum] \sim [\Lambda_{QCD}]\sim [d/dx] \sim [d/dt] =1$$
$$[\vec E]\sim [\vec B] =2$$
and so forth \foot{For practical purposes it is conventional to express
everything in units of energy ($eV$) rather than mass ($gm$).  Thus $m_p\simeq
940$ GeV, $m_e\simeq .511$ MeV,  $1\ F = 10^{-13}\,cm\simeq (200 MeV)^{-1}$,
$1\, K\simeq 10^{-4} \, eV$, etc.}.  Of particular interest to us will be the
dimension of a Lagrange density;  since $\int d^4 x \CL$ is an action --- which
comes in units of $\hbar$ and is dimensionless --- it follows that
$$[\CL] = 4\ .$$

Let us use dimensional analysis to discuss properties of the hydrogen atom.  To
a first approximation, the system is described in terms of one dimensionful
parameter, the electron mass $m_e$, and one dimensionless number, the fine
structure constant $\alpha=1/137$.  If we want to estimate the  size $a_0$ of a
hydrogen atom, since length has dimension $[mass]^{-1}$ it follows that
$a_0\propto m_e^{-1}$, where the proportionality constant is dimensionless.
What is it?  We would guess it is some number of order unity, times some power
of $\alpha$.  Alas, dimensional analysis doesn't tell us the power...we have to
look at the dynamics to realize that the appropriate power is $\alpha^{-1}$.
We arrive at $a_0\simeq 1/(\alpha m_e)$, which in fact is the exact expression
for the Bohr radius.  What about the ground state binding energy $E_0$ of the
hydrogen atom?  $E_0$ has the dimensions of $[mass]$, and dynamics gives us a
proportionality factor of $\alpha^2$, so we estimate $E_0 \simeq \alpha^2 m_e$,
which is in fact only a factor of 2 off from the correct value.

We can go on and ask what wavelength of photon allows one to examine crystal
structure by means of diffraction.  Since atomic sizes are given by $a_0=
1/(\alpha m_e)$, the atomic spacing in a crystal is expected to be similar.
It follows that to see crystal structure, we need photons with wavelength
$\lambda \ltap a_0$, or equivalently energy $E_{\gamma} \gtap \alpha m_e =
\CO(10\, KeV)$ --- visible light won't do, we need X-rays.  On the other hand,
if we wish to estimate the energy of light emitted from an atomic transition in
hydrogen, get $E_{\gamma} \ltap E_0 \simeq \alpha^2 m_e = \CO(10\, eV)$,
corresponding to a wavelength $\lambda_{\gamma} \simeq a_0/\alpha$, several
orders of magnitude larger than the atom itself.

The above analysis may seem  familiar and unimpressive --- after all, while it
is nice that one can easily determine how $m_e$ enters quantities of physical
interest, one also has to keep track of powers of $\alpha$ which involves going
back and examining  the Schr\"odinger equation for hydrogen (which we know how
to solve anyway).
Yet there $is$ something remarkable about the analysis, and it lies in the
sentence:
\item{}
{\it ``To a first approximation, the system is described in terms of one
dimensionful parameter, the electron mass $m_e$, and one dimensionless number,
the fine structure constant $\alpha=1/137$.''}
\lfs
Why should this be true, and what is the approximation we are making?  Why is
the system insensitive to the proton mass? Or the $W$ and $Z$ boson masses?  Or
Newton's constant $G_N$?  Why don't we need to take into account the bottom
quark mass, $m_b \simeq 5$ GeV?   The ratio $r\equiv m_b^2/m_e^2 = 10^{8}$ is a
dimensionless number; couldn't  the ground state  energy of the hydrogen atom
be some function of $r$, namely $E_0 = f(r) \alpha^2 m_e$, where $f(r)$ could
as easily equal $10^{8}$ as $10^{-8}$?

The technique of constructing effective theories allows one to answer such
questions\foot{The concept of effective field theory is mainly associated with
Ken Wilson, although it is an edifice with many architects.  For two quite
dissimilar modern treatments see the reviews by J. Polchinski \ref\pol{J.
Polchinski, Lectures presented at TASI 92, hep-th/9210046} and H. Georgi
\ref\hmg{H. Georgi, Annu. Rev. Nuc. Part. Sci., 43 (1994) 209}.}.  The basic
idea is not to attempt to construct ``a theory of everything'', but to
construct an effective theory that is appropriate to the energy scale of the
experiments one is interested in.  A theory of everything is beyond our
abilities to construct since we cannot probe everything experimentally, and
even if we could, it would contain lots of information extraneous to  any
particular experiment.

Effective field theory techniques get interesting when we wish to look at
effects over a wide range of energies:  then we must understand how effective
theories at different scales are related to each other.  This is useful if one
wishes to relate experiments over a large range of energy scales, or if one has
a theory of high energy physics and wishes to predict the results of low energy
experiments.
In the example of the hydrogen atom and the $b$ quark,   one can show that
$E_0$ depends on the $b$ quark mass in the following way:
$$E_0 = \frac{1}{2} \alpha^2 m_e \(1 +\CO(m_e^2/m_b^2)\)\ .$$
There is a small power law correction to the naive value $\propto
(m_e^2/m_b^2)\sim 10^{-8}$, as well as a hidden dependence of $\alpha$ on the
$b$ quark mass.  When one is only concerned with atomic physics, one can
ignore the $m_b$ dependence of $\alpha$,  since it is already incorporated in
the measured physical value $\alpha=1/137$.  Effective field theories however
allow you to simply compute electromagnetic scattering of electrons at a
center-of-mass energy $\sim 100$ GeV, where one finds that the appropriate
value of for the fine structure ``constant'' is $\alpha\simeq 1/128$ --- a
change due in part to the effects of the $b$ quark.

The outline of these lectures is as follows:
\item{1.} First I discuss how to construct effective theories as an expansion
in operators consistent with low energy symmetries, and how to use dimensional
analysis to extract the interesting physics;
\item{2.} Next I explain how the dimension of the operator determines whether
it is irrelevant, relevant or ``marginal'' to low energy physics;
\item{3.} I then consider ``matching'': how one relates the parameters of a low
energy theory to those of a higher energy theory;
\item{4.}  I then show how quantum corrections can sometimes change the
dimension of an operator and therefore radically change low energy physics.
\item{5.} Finally I mention the application of some of these ideas to the
strong interactions.

\noindent
Each section is followed by some exercises (of widely varying difficulty);  I
encourage you to work through them.
\lfb
\hrule
\lfb
{\bf Exercise 1.} {\sl Estimate the energy scale of rotational excitations of
water in terms of $m_p$, $m_e$ and $\alpha$.   Does your answer explain why
microwaves are used to heat food?}
\lfb
\hrule

\newsec{Dimensional analysis, symmetries, and the separation of scales}

The basic idea behind effective field theories is that a physical process
typified by some energy $E$ can be described in terms of an expansion in
$E/\Lambda_i$, where the $\Lambda_i$ are various physical scales involved in
the problem which have dimension 1 and which are bigger than $E$.  In this
section we show how this simple idea can be incorporated into a predictive
framework.

\subsec{Example 1: Why the sky is blue.}

Consider the question of why the sky is blue. More precisely, consider the
problem of low energy light scattering from neutral atoms in their ground
state, where by ``low energy'' I mean that the photon energy $E_\gamma$  is
much smaller than the excitation energy $\Delta E$ of the atom, which is of
course much smaller than its inverse size or mass:
$$ E_\gamma\ll \Delta E\ll a_0^{-1}\ll  M_{atom}\ .$$
Thus the process is necessarily elastic scattering, and to a good approximation
we can ignore that the atom recoils, treating it as infinitely heavy. Let's
construct an ``effective Lagrangian'' to describe this process.    This means
that we are going to write down a Lagrangian with all interactions describing
elastic photon-atom scattering that are allowed by the symmetries of the world
--- namely Lorentz invariance and gauge invariance.  Photons are described by a
field $A_\mu$ which creates and destroys photons; a gauge invariant object
constructed from $A_\mu$ is the field strength tensor $F_{\mu\nu} =
\partial_\mu A_\nu - \partial_\nu A_\mu$.  The atomic field is defined as
$\phi_v$, where $\phi_v$ destroys an atom with four-velocity $v_\mu$
(satisfying $v_\mu v^\mu=1$, with $v_\mu=(1,0,0,0) $ in the rest-frame of the
atom), while $\phi^{\dagger}_v$ creates an atom with four-velocity $v_\mu$.  So
what is the most general form for $\CL_{eff}$?
Since the atom is electrically neutral, gauge invariance implies that $\phi$
can only be coupled to $F_{\mu\nu}$ and not directly to $A_\mu$.   So
$\CL_{eff}$ is comprised of all local, Hermitian monomials in
$\phi^\dagger_v\phi_v$, $F_{\mu\nu}$,$v_{\mu}$, and $\partial_\mu$.  Certain
combinations we needn't consider for the problem at hand --- for example
$\partial_{\mu}F^{\mu\nu}=0$ for  radiation (by Maxwell's equations); also, if
we define the energy of the atom at rest in it's ground state to be zero, then
$v^\mu\partial_\mu \phi=0$, since $v_{\mu}=(1,0,0,0)$ in the rest frame, where
$\partial_t \phi=0$.  Similarly, $\partial_\mu \partial^\mu \phi = 0$.  Thus we
are led to consider the Lagrangian
\eqn\rayleigh{\eqalign{\CL_{eff} &= c_1 \phi^{\dagger}_v\phi_v
F_{\mu\nu}F^{\mu\nu} + c_2  \phi^{\dagger}_v\phi_v v^\alpha F_{\alpha\mu}
v_\beta F^{\beta\mu}\cr
&+c_3 \phi^{\dagger}_v\phi_v(v^\alpha\partial_\alpha)  F_{\mu\nu}F^{\mu\nu} +
\ldots\cr}}

The above expression involves an infinite number of operators and an infinite
number of unknown coefficients!  Nevertheless, dimensional analysis allows us
to identify the leading contribution to low energy scattering of light by
neutral atoms.  It is straightforward to figure out that
$$[\partial_\mu] = 1\ ,\qquad [F_{\mu \nu}]=2\ ,\qquad [\phi]=\frac{3}{2}\ .$$
The first follows from the fact that $\partial_\mu$ has the dimension of
1/length. The second  is easily determined by noting that the Maxwell
Lagrangian   is $\CL_M = -\frac{1}{4} F^2$, and that $[\CL]=4$.  Finally
$\phi$ is determined by writing a state with no atom as $\ket{0}$, and one atom
as $\ket{A}$, where $\phi^{\dagger}(x)\ket{0}=\Psi_A(x)\ket{A}$, with
$\Psi_A(x)$ being the normalized atomic wavefunction and
$\braket{0}{0}=\braket{A}{A} =1$.   Since $\int d^3x\, |\Psi_A|^2=1$, it
follows that $[\phi]= 3/2$.

Since the effective Lagrangian has dimension 4, the coefficients $c_1$, $c_2$
etc. also have dimensions.  It is easy to see that they all have negative mass
dimensions:
$$[c_1] =[c_2] = -3\ ,\qquad [c_3] = -4$$
and that operators involving higher powers of $\partial\cdot v$ would have
coefficients of even more negative dimension. It is crucial to note that these
dimensions must be made from dimensionful parameters describing the atomic
system --- namely its size $r_0$ and the energy gap $\delta E$ between the
ground state and the excited states. The other dimensionful quantity,
$E_\gamma$, is explicitly represented by the derivatives $\partial_\mu$ acting
on the photon field.  Thus for $E_\gamma\ll \Delta E, r_0^{-1}$ the dominant
effect is going to be from the operator in $\CL_{eff}$ which has the {\it
lowest} dimension.  There are in fact two leading operators, the first two in
eq. \rayleigh, both of dimension 7.  Thus low energy scattering is dominated by
these two operators, and we need only compute $c_1$ and $c_2$.

What are the sizes of the coefficients?  To do a careful analysis one needs to
go back to the full Hamiltonian for the atom in question interacting with
light, and ``match'' the full theory to the effective theory.  We will discuss
this process of matching later, but for now we will just estimate the sizes of
the $c_i$ coefficients.  We first note that   extremely low energy photons
cannot probe the internal structure of the atom, and so the cross-section ought
to be classical, only depending on the size of the scatterer.  Since such low
energy scattering can be described entirely in terms of the coefficients $c_1$
and $c_2$, we conclude that
$$c_1 \simeq c_2\simeq r_0^3\ .$$
The effective Lagrangian for low energy scattering of light is therefore
\eqn\rayleighii{\CL_{eff} = r_0^3 \( a_1 \phi^{\dagger}_v\phi_v
F_{\mu\nu}F^{\mu\nu} + a_2  \phi^{\dagger}_v\phi_v v^\alpha F_{\alpha\mu}
v_\beta F^{\beta\mu}\)}
where $a_1$ and $a_2$ are dimensionless, and expected to be $\CO(1)$.  The
cross-section (which goes as the amplitude squared) must therefore be
proportional to $r_0^6$.  But a cross section $\sigma$ has dimensions of area,
or $[\sigma]=-2$, while $[r_0^6]= -6$.  Therefore the cross section must be
proportional to
\eqn\rcross{\sigma\propto E_\gamma^4 r_0^6 \ ,}
growing like the fourth power of the photon energy.  Thus blue light is
scattered more strongly than red, and the sky looks blue.

Is the expression \rcross\ valid for arbitrarily high energy?  No, because we
left out terms in the effective Lagrangian we used.  To understand the size of
corrections to \rcross\ we need to know the size of the $c_3$ operator (and the
rest we ignored).  Since $[c_3]=-4$, we expect the effect of the $c_3$ operator
on the scattering amplitude to be smaller than the leading effects by a factor
of $E_\gamma/\Lambda$, where $\Lambda$ is some energy scale.  But does
$\Lambda$ equal $M_{atom}$, $r_0^{-1}\sim \alpha m_e$ or $\Delta E\sim \alpha^2
m_e$?   The latter is the smallest scale and hence the most important.  We
expect our approximations to break down as $E_\gamma \to \Delta E$ since for
such energies the photon can excite the atom.  Hence we predict
\eqn\rcrossii{\sigma\propto E_\gamma^4 r_0^6\(1 +\CO(E_\gamma/\Delta E)\) .}
The Rayleigh scattering formula ought to work pretty well for blue light, but
not very far into the ultraviolet.  Note that eq. \rcrossii\ contains a lot of
physics even though we did very little work.  More work is needed to compute
the constant of proportionality.

\subsec{ Example 2: The binding energy of charmonium to nuclei.}

Closely related to the above example is the calculation of the binding energy
of charmonium (a $\bar c c$ bound state, where $c$ is the charm quark) to
nuclei.  In the limit that the charm quark mass $m_c$ is very heavy, the
charmonium meson can be thought of as a Coulomb bound state, with size $\sim
\alpha_s(m_c) m_c$, where $\alpha_s(m_c)$ is a small number (more on this
later).   When inserted in a nucleus, it will interact with the nucleons by
exchanging gluons with nearby quarks.  Typical momenta for gluons in a nucleus
is set by the QCD scale $\Lambda_{QCD}\simeq 200$ MeV.  For large $m_c$ then,
the wavelength of gluons will be much larger than the size of the charmonium
meson, and so the relevant interaction is the gluon-charmonium analogue of
photon-atom scattering considered above.  The effective Lagrangian is just
given by \rayleighii, where $\phi$ now destroys charmonium mesons, and
$F_{\mu\nu}$ is replaced by $G_{\mu\nu}^a$, the field strength for gluons of
type $a=1,\ldots,8$.  The coefficients $a_{1,2}$ may be computed from QCD.  To
compute the binding energy of charmonium we need to compute the matrix element
$$\bra{N,\bar c c} \int d^3 x\,\phi^{\dagger}\phi G^a_{\mu\nu}
G^{a\,\mu\nu}\ket{N,\bar c c} $$
(as well as the matrix element of the other operator in \rayleighii), which we
do not know how to do precisely since the system is strongly interacting.  We
can estimate its size by dimensional analysis though, getting
$$E_B \sim r_0 ^3 \Lambda_{QCD}^4 \simeq {\Lambda_{QCD}^4  \over (\alpha_s
m_c)^3}\ .$$
This problem is discussed with greater sophistication in ref. \ref\charm{M.
Luke, A. Manohar, M. Savage, Phys. Lett. B288 (1992) 355.}.

\subsec{ Example 3: The cross section for low energy neutrino interactions.}

The term ``weak interactions'' refers in general to any interaction mediated by
the $W$ or $Z$ bosons, whose masses are $80$ GeV and $91$ GeV respectively.
Since their couplings are rather weak, it is usually a decent approximation to
only consider first order perturbation theory, namely Feynman graphs fig. 1(a).

\topinsert
\centerline{\epsfysize=1in€\epsfbox{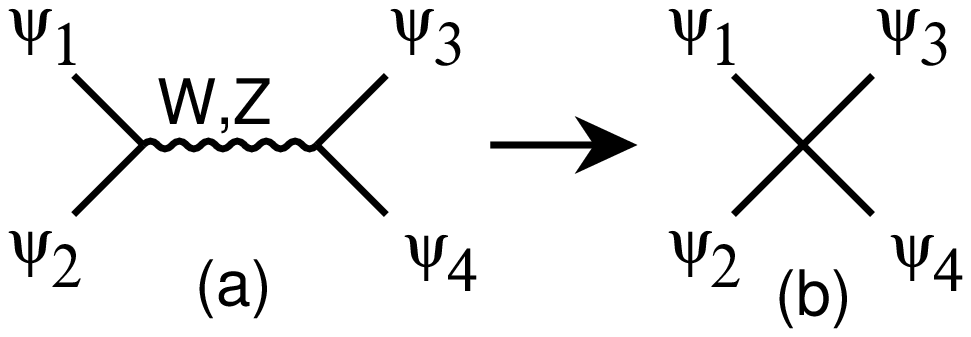}}
\caption{Fig. 1. (a) Tree level $W$ and $Z$ exchange between four fermions. (b)
The effective vertex in the low energy effective theory (Fermi interaction).}
\endinsert

These interactions describe $2\to 2$ scattering of fermions, or $1\to 3$
decays. The $W$ and $Z$ propagators (in a particular choice of gauge) are given
by $-i g_{\mu\nu}/(q^2-M^2)$, where $q$ is the four-momentum transferred.  For
low energy processes, $q^2 \ll M^2$ and one never has enough energy to make a
physical $W$ or $Z$, so there is no reason to include them in the theory.  Thus
the low energy effective theory just has  the contact interactions shown in
figure 1b:
\eqn\clw{\CL_{weak} \sim G\psi_1\psi_2\psi_3\psi_4,}
where $\psi_i$ represent fermion fields (for either quarks or leptons).
Since the Lagrangian for a noninteracting fermion is $\CL_f = \bar \psi (i\dsl
-m)\psi$, it follows that
$$[\psi] = \frac{3}{2}\ ,$$
and so the coupling $G$ in eq.\clw\ has dimension $[G]= -2$.  You can estimate
its size by equating the processes fig 1a and fig. 1b and it is roughly given
by $g^2/M^2$, where $g$ and $M$ are the dimensionless coupling constant and
mass of the $W$ or $Z$.  (This is ``matching'', and you will do this more
precisely in a later exercise).

Since neutrinos only interact through the weak force, it follows that low
energy  neutrinos ($E_\nu \ll M_W$) interact with matter through an operator of
the form \clw, where two of the $\psi$'s are neutrino fields, and the other two
are either quark or lepton fields.  Thus the neutrino cross-section $\sigma$,
which has dimension -2,  must be proportional to $G^2$ which has dimension -4.
Therefore the cross-section must scale with energy as
\eqn\snu{\sigma_\nu \simeq G^2 s}
for low energy neutrinos, where $s $ equals the square of the total energy in
the center of momentum frame.

\lfb\hrule
\lfb
{\bf Exercise 2.} {\sl Use the effective Lagrangian to explain why the force
between to static neutral atoms at a separation $R\gg a_0$ scales like $1/R^7$.
You should be able to get this from dimensional analysis of the two photon
exchange process.  Can you explain why there isn't any contribution from one
photon exchange due to the operator $\phi_v^{\dagger} i\darr \partial_\mu
\phi_v v_\nu F^{\mu\nu}$?  Can you explain why the approximations made in the
effective field theory are expected to be invalid for $R\ltap a_0/\alpha$? For
a detailed discussion of why one finds $1/R^7$ instead of
the nonrelativistic result $1/R^6$, see ref. \ref\izub{C. Itzykson, J.- B.
Zuber, {\it Quantum Field Theory}, McGraw Hill Inc., NY (1980), pp. 141, 365.}.
}
\lfb\hrule
\lfb
{\bf Exercise 3.} {\sl The $\mu$ and the $\tau$ have the same weak
interactions, and so the amplitudes for decay via $W$ exchange  $  \mu \to e
\bar\nu_e\nu_\mu$ and $ \tau \to e \bar\nu_e\nu_\tau$ are equal.   Since the
$\tau$ is heavier, it has more ways to decay than the $\mu$. The mass and
lifetimes of the two particles  are
$$m_\mu = 106\ MeV\ ,\qquad T_\mu = 2.2\times 10^{-6}\ sec.,$$
$$m_\tau = 1777\  MeV\ ,\qquad T_\tau = 3.0\times 10^{-13}\  sec. $$
Given that the $\mu$ decays 100\% of the time via $\mu\to  e \bar\nu_e\nu_\mu$,
calculate the fraction of $\tau$ decays which are of the form   $\tau\to  e
\bar\nu_e\nu_\mu$. All you need to know is that $[G]=-2$ in eq. \clw.  How does
your answer compare with the observed branching ratio  $BR_{\tau\to  e
\bar\nu_e\nu_\tau} = 18.01\pm 0.18\%$?}

\lfb
\hrule\lfb
{\bf Exercise 4.} {\sl The partial mean lifetime of the proton in the decay
$p\to e^+\pi^0$ is known to be greater than $1.3\times 10^{32}$ years.  Suppose
that new physics at a scale $\Lambda$ does give rise to this decay (for
example, through the tree level exchange of a particle with mass $\Lambda$,
analogous to the interaction in fig. 1).  What is an approximate lower bound on
 $\Lambda$ ?  (Hint: Find the lowest dimension operators made up of quark and
lepton fields that could give rise to this decay mode).  }
\lfb
\hrule\lfb
{\bf Exercise 5.} {\sl Suppose that there are $\Delta B=2$ baryon violating
operators due to new physics at a scale $\Lambda$, but no $\Delta B=1$
operators, so that the proton is stable, but $n-\bar n$ oscillations can occur.
 Such oscillations have not been seen, and the lower bound on the oscillation
rate is $1.2\times 10^8$ sec. How does this translate into a bound on the scale
$\Lambda$?  }
\lfb\hrule\lfb
{\bf Exercise 6.} {\sl Estimate the cross-section for photon-photon scattering
at energies well below the electron mass, $E_\gamma\ll m_e$.  Since $\alpha =
1/137$, counting powers of $\alpha$ matters! }
\lfb\hrule\lfb

\newsec{The relevant, the irrelevant, and the marginal}

So far I have only discussed examples where the operator has dimension greater
than 4, so that the coefficient has negative dimension and the resulting
cross-section or decay width therefore becomes smaller as the energy scale $E$
of the interaction gets smaller.  Even though these are often the most
interesting interactions --- since they are harbingers of new physics at
energies well above $E$ --- these sorts of interactions are called {\it
irrelevant}.  The rationale is that at low energies, their effects are small
(for example, see eqs. \rcrossii, \snu.).    In contrast, operators with
dimension less than 4, whose coefficients have positive dimension, are called
{\it relevant} operators because they become more relevant at lower $E$.
Ignoring quantum corrections, the only relevant operators one can write down in
a relativistic field theory in four dimensions are
\item{$\bullet$} The unit operator (whose coefficient is the cosmological
constant) which is dimension 0;  \item{$\bullet$} Boson   mass terms, which are
dimension 2;
\item{$\bullet$} Fermion mass terms, which are dimension 3;
\item{$\bullet$} 3-scalar ($\phi^3$) interactions, also dimension 3.
\lfm
(Terms linear in a scalar field can be removed by shifting its value).

An example is the electron mass, arising from the dimension 3 operator $\bar
\psi \psi$ with coefficient $m_e$.  In high energy scattering ($E_e\gg m_e$)
the effects of the electron mass are negligible.  However, the effects of the
electron mass are very important at energies comparable to $m_e$.  In fact,
exercise 3. is only simple if one not only assumes that the momentum scales in
$\mu$ and $\tau$ decay are low compared to $M_W$, but also that they are high
compared to $m_e$, so that one could ignore the electron mass.  As another
example, consider two real scalar fields $\phi$ and $\Phi$ with a Lagrangian of
the form
\eqn\iiis{\CL = \half (\partial \phi)^2 + \half (\partial \Phi)^2 - \half
{m^2} \phi^2 - \half {M^2}  \Phi^2 - \half \kappa \phi^2\Phi\ .}
We will assume
$$m\simeq \kappa\ll M\ .$$
We can see that unlike fermion fields, scalar fields have dimension 1, which
means that the coupling $\kappa$ does as well:
$$[\phi] = [\Phi] = [\kappa] = 1\ .$$
By our definition above, the three scalar interaction is {\it relevant}.
Consider $\phi\phi \to \phi\phi$ scattering at tree level in this model.  First
take the case where the center of mass energy $E_\phi$ is much greater than
$m$, $M$, and $\kappa$.  Then the scattering amplitude from a graph like fig.
1a --- with the $\psi_i$ replaced by $\phi$ and the $W,Z$ replaced by $\Phi$
--- is proportional to $\kappa^2$ and the cross section must go as
$$\sigma_{2\phi\to 2\phi}\vert_{E_\phi \gg m,M,\kappa} \propto  ({\kappa\over
E_\phi})^4 {1\over E_\phi^2} $$
which goes rapidly to zero for large $E_\phi$.  Now look at the scattering
cross section at an energy $E_\phi$ satisfying $m\ll E_\phi \ll\kappa, M$, so
that the $\phi$ particles are still relativistic, but the $\Phi$ propagator and
be contracted to a point as in figure 1b.  Now the cross section goes as
$$\sigma_{2\phi\to 2\phi}\vert_{m\ll E_\phi \ll\kappa, M} \propto
({\kappa\over M})^4 {1\over E_\phi^2}$$
Contrasting this low energy cross section with that for neutrinos in \S 2.3
explains why  $\kappa$ interaction is said to be relevant at low energies,
while Fermi interaction is called irrelevant.

Operators with dimension 4 lie between relevancy and irrelevancy and are called
{\it marginal}.  Examples of marginal interactions are
\item{$\bullet$} $\phi^4$ interactions;
\item{$\bullet$} Yukawa interactions ($\bar\psi \psi \phi$);
\item{$\bullet$} Gauge interactions (interactions of a gauge boson with itself,
a scalar, or a fermion).
\lfm
As we will see, marginality is an insecure position to be in, and quantum
corrections will almost always change such operators from marginal to either
relevant or irrelevant.

In each of the examples in the previous section we focussed on irrelevant
interactions.  The only reason why this was interesting was that in each case,
irrelevant operators gave the leading contribution to the process... and
because they weren't {\it too} irrelevant.  For example, neutrinos {\it only}
interact with matter through irrelevant operators...so if one sees any evidence
of low energy neutrino scattering, one is seeing irrelevant operators.  In
contrast,  $e^+ e^-$ scattering has an electromagnetic contribution from photon
exchange.  Since the photon-electron coupling is a marginal operator, at low
energies electromagnetic interactions dominate the weak interaction
contribution.  (No
coincidence that these are called weak interactions!).   Now imagine a world
where the $W$ and $Z$ masses were $10^{16}$ GeV.  In this world there would be
practically no discernible weak interaction effects.  The neutron would have a
lifetime greater than $10^{30}$ years, and there would be no radioactivity; no
one would have guessed that the neutrino existed, because it would not interact
with anything.  All we would discern in particle collisions and spectra would
be the strong and electromagnetic interactions.

In fact, in any situation where there is a large gap between the energy where
one is doing experiments and the energy scale of new physics, the effective
theory one  constructs will only consist of marginal and relevant operators...
such theories are called ``renormalizable'' and are a natural outcome when
there is a large hierarchy of physical scales. This typically results in a vast
simplification of the physics one needs to consider, as seen in the next
example.

\subsec{ Example: the success of Landau liquid theory.}

A condensed matter system can be a very complicated environment; there may be
various types of ions arranged in some crystalline array, where each ion has a
complicated electron shell structure and interactions with neighboring ions
that allow electrons to wander around the lattice.  Nevertheless, the low
energy excitation spectrum for many diverse systems can be described pretty
well as a ``Landau liquid'', whose excitations are fermions with some
complicated dispersion relation but no interactions.  Why this is the case can
be simply understood in terms of effective field theories, modifying the
dimension counting used above to suit a nonrelativistic system with a Fermi
surface \foot{The treatment here follows that of Polchinski in ref. \pol.}.

Let us assume that the low energy spectrum of the condensed matter system has
fermionic excitations with arbitrary interactions above a Fermi surface
characterized by the fermi energy $\epsilon_F$ ; call them
``quasi-particles''. Ignoring interactions, the action can be written as
\eqn\sfree{S_{free}=\int dt\,\int d^3p\, \sum_{s=\pm\half} \[
\psi_s(p)^{\dagger} i\partial_t \psi_s(p) - (\epsilon(p)-\epsilon_F)
\psi^{\dagger}_s(p)\psi_s(p)\]}
where an arbitrary dispersion relation $\epsilon(p)$ has been assumed.  Now let
us consider higher dimension operators...but how should we count ``dimension''?
 In the relativistic case, we defined mass dimension in a simple way, since we
wanted to do an expansion in $E/\Lambda$, where $E$ was the scale of the
experiment and $\Lambda$ was a physical scale associated with the system being
probed.   In a nonrelativistic system we   identify the scaling dimension
with momentum, in which case energy scales like $p^2$.  Furthermore, it doesn't
make sense to expand around $p=0$ since an excitation cannot have a momentum
vector inside the Fermi surface.   So we write the momentum as
\topinsert
\centerline{\epsfysize=1in€\epsfbox{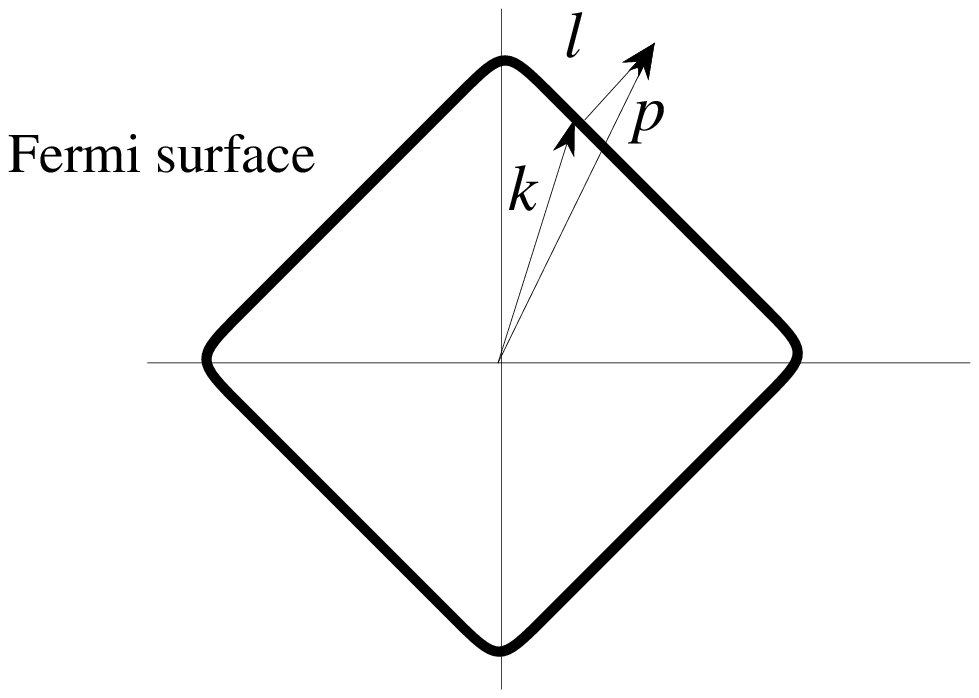}}
\caption{Fig. 2. The momentum $\vec p$ of an excitation above the Fermi surface
is divided into a component $\vec k$ on the Fermi surface, and a component
$\vec\ell$ perpendicular to the surface.  The length of $|\vec \ell|$ is the
quantity one wants to scale.}
\endinsert
$$\vec p = \vec k+ \vec \ell$$
where $\vec k$ lies on the Fermi surface and $\vec \ell$ is perpendicular to it
(fig. 2).  Then $\vec \ell$ is the quantity we vary in experiments and so we
define the dimension of operators by how they must scale so that the theory is
unchanged when we change $\vec \ell \to r\vec \ell$.  If an object scales as
$r^n$, then we say it has dimension $n$.  Then  $[k]=0$, $[\ell]=1$, and $[\int
d^3p = \int d^2k d\ell] = 1$. And if we define the Fermi velocity as
$\vec\nabla_p\epsilon$, then for   $\ell\ll k$,
$$\epsilon(\vec p) - \epsilon_F = \vec\ell\cdot \vec v_F(\vec k) +\CO(\ell^2)\
,$$
and so $[\epsilon - \epsilon_f] = 1$ and $[\partial_t] = 1$. Given that the
action
\sfree\ isn't supposed to change under this scaling,
$$[\psi] = -\half\ .$$

Now consider   an interaction of the form
$$S_{int} = \int dt\, \int \prod_{i=1}^4 (d^2 k_i d\ell_i) \delta^3(\vec
P_{tot}) C(k_1,\ldots,k_4)\psi_s^\dagger(p_1) \psi_{s}(p_2)
\psi_{s'}^{\dagger}(p_3)\psi_{s'}(p_4)\ .
$$
This will be relevant, marginal or irrelevant depending on the dimension of
$C$. Apparently $[\delta^3(P_{tot}) C] = -1$.  So how does the $\delta$
function
scale?  For generic $\vec k$ vectors, $\delta(\vec P_{tot})$ is a constraint on
the $\vec k$ vectors that doesn't change much as one changes $\ell$, so that
$[\delta^3(\vec P_{tot})]=0$.   It follows that $[C]=-1$ and that the four
fermion interaction is irrelevant...and that the system is adequately described
in terms of free fermions (with an arbitrarily screwy dispersion relation).
This effect is known in nuclear physics, where Pauli blocking allows a strongly
interacting system of nucleons to have single particle excitations.

It is amusing that when a pair of $\vec k_i$ vectors are within $\CO(\ell)$ of
cancelling each other, then the scaling dimension of the delta function changes
from 0 to $-1$.  To see this, fix set the $\ell$'s to zero, and fix the
incoming momenta $\vec k_1$ and $\vec k_2$.  The $\delta$-function then
generically constrains three out of the four degrees of freedom in the outgoing
momenta $\vec k_3$ and $\vec k_4$ in terms of $\vec k_1+\vec k_2$.  However, if
$\vec k_1+\vec k_2=0$, then
$\vec k_3+\vec k_4$ must equal zero, but that only constrains two of the four
degrees of freedom (assuming a parity symmetric Fermi surface).  Therefore the
delta function $\delta^3(p)$ must scale like
   $\delta^2(k)\delta(\ell)$, and so for these head-on collisions between
particles at opposite sides of the Fermi sea, $[C]=0$, and the interaction is
marginal.   Quantum corrections either make it either irrelevant of relevant;
it turns out that for $C$ attractive, the interaction becomes relevant, and if
it is repulsive it becomes irrelevant.  In the former case, the interaction
between such quasiparticles becomes strong near the Fermi surface, and can lead
to pairing and superconductivity.  See ref. \pol\ for more about this.

\lfb\goodbreak\hrule
\lfb
{\bf Exercise 7.} {\sl How would you couple phonons to the fermions in a Landau
liquid?  Would the phonon - fermion coupling be relevant, irrelevant, or
marginal?}
\lfb
\hrule

\newsec{Quantum corrections and renormalization}

It is fine to call a higher dimension operator irrelevant when one is computing
amplitudes at tree level, and the momenta flowing through the vertices is
small.  But what happens when one calculates quantum corrections (loop graphs)
involving these irrelevant interactions and   integrates over intermediate
states of all energies? Do the irrelevant operators become important? A field
theory with irrelevant operators used to fill field theorists with horror,
since they were ``nonrenormalizable''.  This meant that rather than having a
finite number of counterterms that had to be fixed by some experimental
measurement, one needed an infinite number.  Such theories were thought to be
unpredictive.  QED is a good example of a renormalizable theory:  Only two
measurements are needed to fix the counterterms, namely $\alpha$ and $m_e$.
Once these quantities are measured in one set of experiments, all other QED
processes can be predicted.  In  a theory with irrelevant operators, however,
extra insertions of the operator in a graph makes it more divergent.  In a
theory with a Fermi interaction, for example  --- $(\bar\psi \psi)^2$ ---  one
finds one needs counterterms for all $(\bar\psi \psi)^{2n}$ operators.
Furthermore, these operators can in general renormalize relevant operators,
such as the fermion mass, so it seems that all of these infinite number of
interactions must be fit to experiment and nothing can be predicted.

This quandary is avoided if one uses a mass independent renormalization scheme
(dimensional regularization), and thinks of the effective theory not as an
expansion in operators, but as an expansion in inverse powers of some large
physical scale $\Lambda$.  Let us assume that we wish to do experiments at some
momentum scale $p$ and that the relevant operators have coefficients set by a
scale $m\ltap p$.  In contrast, the irrelevant operators have coefficients
which are inverse powers of $\Lambda\gg m,p$.  For example, a theory of a
fermion with mass $m$ and higher order interactions:
$$\CL = \bar\psi i\dsl\psi - m\bar\psi\psi - {a\over \Lambda^2} (\bar\psi
\psi)^2 -
{b\over \Lambda^4} (\bar\psi \psi)^3 - \ldots$$
Now consider the divergent graph in fig. 3.
\topinsert
\centerline{\epsfysize .7in\epsfbox{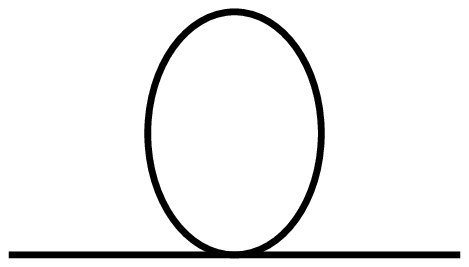}}
\caption{Fig. 3. A divergent one-loop radiative correction to the fermion mass
and kinetic term in a theory with a $(\bar\psi\psi)^2$ interaction.}
\endinsert
This graph gives  a divergent contribution to the mass operator $\bar\psi\psi$
proportional to
$${i\over \Lambda^2} \int {d^4 q\over (2\pi)^4} {m\over q^2- m^2}\ .$$
When Wick rotated into Euclidian space and defined by dimensional
regularization, the above integral equals (see eq. (A.1) in the Appendix)
$${ m^3\over 16\pi^2 \Lambda^2}\( -{1\over \epsilon} +\gamma - 1
+\ln\[{m^2\over 4 \pi \mu^2}\]\)$$
where we are in $4-2\epsilon$ dimensions, and $\mu$ is the renormalization
scale that creeps into the problem\foot{For a discussion of dimensional
regularization, see for example refs. \ref\collins{J. C. Collins, {\it
Renormalization}, Cambridge University Press, Cambridge, 1984}, \ref\ramond{P.
Ramond, {\it Field Theory, A Modern Primer}, Benjamin/Cummings, Reading MA,
1981.}.  For those familiar with the concepts, some useful formulas are
included as an appendix to these lecture notes.}.  In a mass independent
subtraction scheme  we put in a one-loop counterterm that cancels the infinite
part of this graph, as well as a mass independent finite part.  For example, in
the $\bar{MS}$ scheme, we subtract the part proportional to
$${  a m^3\over 16\pi^2 \Lambda^2}\( -{1\over \epsilon} +\gamma - 1 -\ln 4 \pi
\)\ .$$
We are left with a finite contribution to the fermion mass equal to (up to an
$\CO(1)$ numerical factor which I have dropped)
\eqn\dmass{\delta m \sim {a m^3\over 16\pi^2 \Lambda^2} \ln\[{m^2\over
\mu^2}\] \ .}
We choose a convenient scale $\mu$ and fit $(m+\delta m)$ to experiment (once
one has also calculated the one-loop wave function renormalization).

The important point I wish to make is that
$${\delta m\over m} \propto {m^2\over 16\pi^2 \Lambda^2}\ ;$$
it is {\it small} --- it needs to be taken into account when probing effects
proportional to $1\over\Lambda^2$, but not otherwise.  Note that this would not
have been the case if we had simply taken $\Lambda$ to be a physical momentum
cutoff and not renormalized...then, since the fermion loop graph is
quadratically divergent, we would have found $\delta m \sim (m/\Lambda^2)\times
\Lambda^2 \sim m$.  This would be a ludicrous state of affairs --- we would
have to understand quantum gravity, for example, to compute radiative
corrections to $e^+e^-$ scattering.

The above example has several important features which I wish to draw your
attention to:
(i) The correction to the electron mass $\delta m$ is suppressed by
$m^2/\Lambda^2$;
(ii) $\delta m$ has a logarithmic dependence on the fermion mass and the
renormalization scale $\mu$;
(iii)  The corrections to the fermion mass are proportional to the fermion
mass.  Each of these three points is worth commenting on:
\subsec{The size of radiative corrections}
Concerning the first point: it is an obvious and general result that in a mass
independent subtraction scheme, corrections to low dimension operators due to
high dimension operators are always suppressed by powers of $p/\Lambda$ and
$m/\Lambda$.  This is {\it not} what one would finds simply putting $\Lambda$
in as a momentum cutoff for one's integrals.  It is an obvious result because
the only new mass scale induced by dimensional regularization is $\mu$, and
that can be seen to only enter logarithms.  Thus an integral with dimension $n$
will be proportional to the $n^{th}$ power of the physical scales in the
problem $p$ and/or $m$.  The scale $\Lambda$ only enters the problem raised to
negative powers at the vertices.  Thus the graph is always proportional to
$(p/\Lambda)^n$ where $n$ is the combined powers from the vertices.  No
positive power of $\Lambda$ is generated by the loop integral.

The fermion mass in our theory {\it does}  receive an infinite number of
corrections from the infinite number of higher dimension operators, and  they
are only computable  if I measure all of the coefficients of these operators.
However the theory remains predictive, since at any finite order in
$m/\Lambda\ll 1$ there are a finite number of contributions to $\delta m$.

\subsec{Radiative logarithms and the scale $\mu$}
The renormalization scale $\mu$ enters through logarithms of $\mu/m$ or
$\mu/p$.  If we could sum up all orders in perturbation theory, all our answers
would be $\mu$ independent.  However, we stop at finite order, and our choice
of $\mu$ can affect how quickly the perturbative expansion converges, since
higher loop graphs yield higher powers of $\ln\mu^2/p^2$.  Thus we should
optimize perturbation theory by choosing $\mu$ to minimize the logarithm.  When
comparing experiments at widely different physical scales, we may run across
large logarithms then of the form $\ln (p_1^2/ p_2^2)$ since the same $\mu$
cannot make the logs in the two processes simultaneously small.  These large
logs can be resummed using the renormalization group, discussed in a later
section.

\subsec{Symmetry and naturalness}
I noted that  $\delta m \propto m$ in eq. \dmass.  This is because $m\to 0$
increases the
symmetry of the theory:  in the above example the symmetry $\psi \to \gamma_5
\psi$, $\bar\psi \to -\bar\psi\gamma_5 $ is a symmetry of the Fermi interaction
(and kinetic term), but not the mass term.  If $m=0$ it follows that $\delta m$
must also vanish. Therefore it is {\it natural} that the fermion mass might be
small compared to other physical scales in the problem.  In contrast, a scalar
mass term $\phi\phi^*$ does not usually break a symmetry --- the only
exceptions are if the theory is supersymmetric, or if the scalar is a Goldstone
boson.  The latter is important for pion physics and  chiral perturbation
theory.  Even if the tree level scalar mass is
zero, it will get radiatively corrected by other fields it couples to. Thus it
is {\it unnatural} for there to be a light scalar coupled to high energy
fields.  Since scalars presumably couple to gravity, typified by the Planck
scale $m_P = 10^{19}$ GeV, one has to wonder why the Higgs boson in the
standard model has a mass in the $10^2$ to $10^3$ GeV mass range. (It has been
suggested that in fact either there is no scalar Higgs boson, or that it is a
Goldstone boson, or that it is a member of a supersymmetric multiplet of
particles).

It is ironic that it used to be that people were worried about theories with
irrelevant operators being sick.  In fact what we see is that irrelevant
operators cause no problems; it is the relevant operators that we must worry
about.  If relevant operators appear in the effective field theory, then they
must be set by a scale much less than $\Lambda$ (else they wouldn't be in the
effective theory below $\Lambda$).  But if their coefficients are much smaller
than $\Lambda$ without a symmetry reason, then we are baffled.  The prime
example is the cosmological constant, namely the dimension 4 coefficient of the
operator {\bf 1}, otherwise known as the vacuum energy density.  There is no
known symmetry that appears relevant to our world that is increased by setting
the vacuum energy density to zero, yet from cosmological observations, the
vacuum energy is known to be $\ltap 10^{-46}\ GeV^4$ \ref\kt{E. Kolb, M.
Turner, {\it The Early Universe}, Addison Wesley, Redwood City CA, 1990.}.  The
smallness of the cosmological constant should be taken as a warning: it appears
contrary to effective field theory dogma, so the dogma may be flawed.

\lfb\hrule
\lfb
{\bf Exercise 8.} {\sl Compute both the wavefunction and mass corrections from
the graph in fig. 3, using the $\bar{MS}$ scheme. See the Appendix for
dimensional regularization formulas. }
\lfb
\hrule

\newsec{Matching}

Consider doing experiments with photons and electrons entirely within the
context of QED.  There are three different regimes for scattering experiments
which one might consider:
\item{1.} Either photons  or electrons or both in the incoming state,   with
momentum transfer large compared to $m_e$;
\item{2.}    electrons in the incoming state, but at momentum transfer much
smaller than $m_e$;
\item{3.}  photons in the incoming state, but momentum transfer much less than
$m_e$.

\lfm
In the first case one needs to compute the relevant amplitude in the full QED
theory, although at high energy  one might make the approximation that the
electron is massless.  Furthermore, the fine structure constant has to be
adjusted from its low energy value $\alpha = 1/137$, and effect due to quantum
corrections which we discuss in a later section.  The second case  is a little
funny --- we can ignore much of the complexity of QED since we do not have
enough energy to produce positron-electron pairs, yet we still need to include
both electrons and photons in the theory;  I briefly mention the techniques one
uses in this case in \S7.   For the third case one need only consider an
effective theory with
photons...why include electrons if one never sees any?

The low energy theory of photons alone looks like
$$\CL_{eff} = -\frac{1}{4} F_{\mu\nu}F^{\mu\nu} + {1\over m_e^4}\(a
(F_{\mu\nu}F^{\mu\nu})^2 + b(F_{\mu\nu}\tilde F^{\mu\nu})^2 \) + \CO(1/m_e^8)\
, $$
the most general local, hermitian theory invariant under Lorentz, gauge, charge
conjugation and
parity transformations. (Can you show why there are no irrelevant operators of
dimension 6?).  This is not QED because it distorts high energy physics...but
we do care that it correctly reproduces low energy phenomenology If we did not
know about QED, we could treat this as a
phenomenological theory and try to fit $a$ and $b$ to measured scattering cross
sections.  However, we do know QED, and so we can compute $a$ and $b$.  To do
this we simply require that $\CL_{QED}$ and $\CL_{eff}$ give us the same
physical predictions at low energy. In general, ensuring that the effective
theory agrees in
its predictions with the full theory to any desired order of accuracy is called
``matching''.  What we are matching is the value of Green's functions  in the
two theories.  Effective field theories are designed to reproduce all of the
infrared (light particle) physics of the full theory, while distorting the high
energy behavior to make calculations simpler.  All of the interesting infrared
effects in the full theory due to light particles are explicitly included; only
the effects of the heavy particles or high energy modes must be mocked up.  So
the correct thing to do is match all the ``one  light particle irreducible''
(1LPI) diagrams (diagrams that do not fall apart when one light particle line
is cut), since these are the graphs that contain either a heavy particle, or
high energy modes of a light particle.   We cannot do this exactly of course,
but we can do it systematically in a ``loop'' expansion, which is an expansion
in powers of the numbers of loops in a diagram, or equivalently, powers of
$\hbar$ \foot{It may seem funny expanding in a dimensionful quantity we set to
unity!  However the loop expansion can be seen to be  consistent with a
perturbative expansion in coupling constants   --- see Coleman's lecture
``Secret Symmetries'' in ref. \ref\col{S. Coleman, Aspects of Symmetry}.}.

\subsec{Example: the $\phi^2\Phi$ interaction.}
Rather than discussing QED, I will consider a toy model that exhibits nicely
the matching procedure.  It is the theory in eq. \iiis\  with a light scalar
$\phi$ coupled to a heavy scalar $\Phi$ via the interaction
$\half\kappa\phi^2\Phi$.  (Never mind that the vacuum energy is unbounded
below; one won't see this in perturbation theory).  Suppose we are interested
in $2\phi\to2\phi$ scattering at energies much below the $\Phi$ mass $M$.  The
graphs we have to match to order $\hbar$ are those in fig. 4.
\topinsert
\centerline{\epsfysize 2.5in\epsfbox{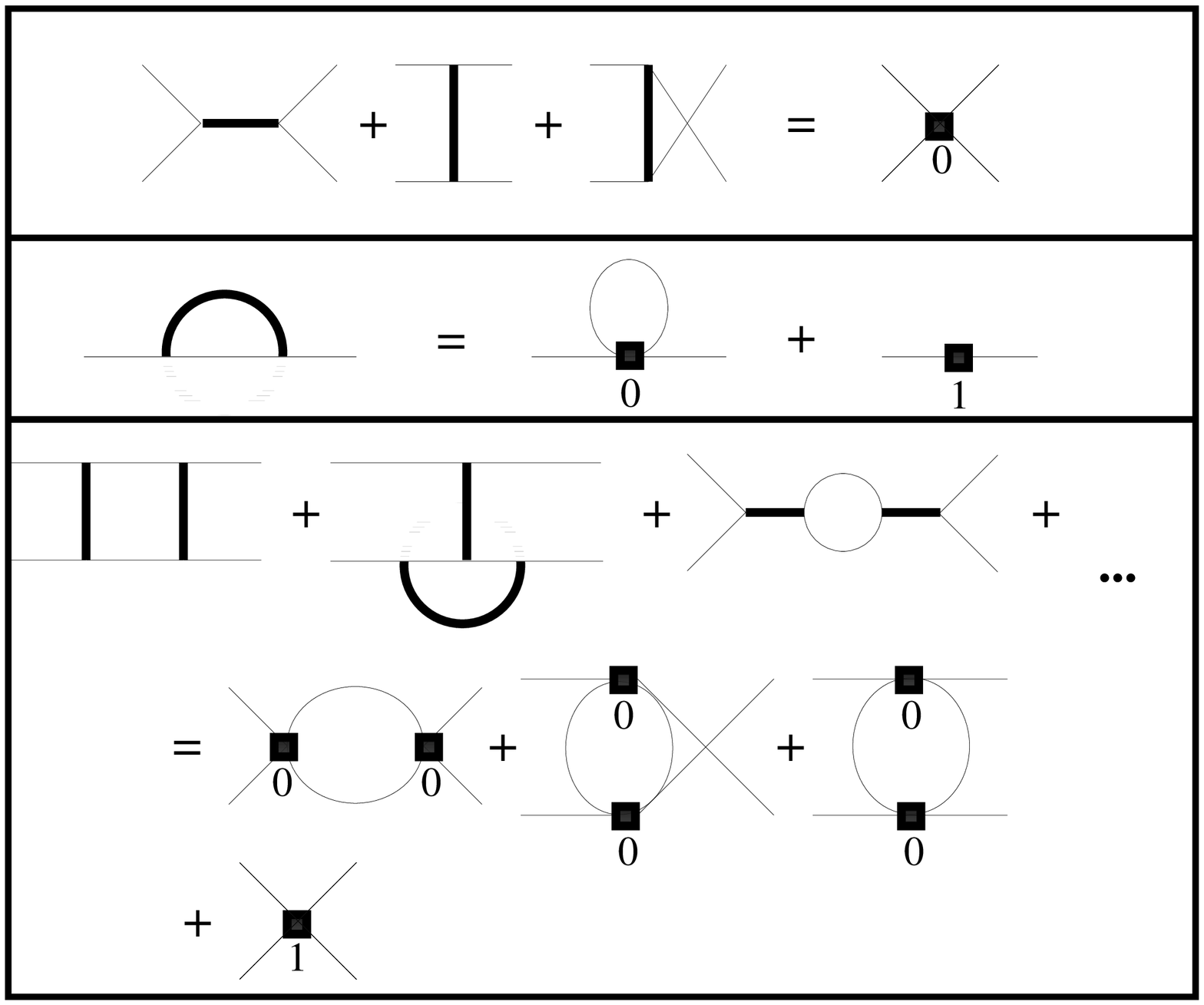}}
\caption{Fig. 4. Matching  conditions for the theory of eq. \iiis.  Diagrams on
the left are in the full theory, while those on the right are in the effective
theory. Heavy lines correspond to the heavy scalar propagator; numbers beneath
the vertices count the loop order of the matching condition.  The first row is
the complete tree level matching condition; second and third rows are the
one-loop matching conditions for the two- and four-point vertices respectively.
  Note that
matching conditions are {\it not} simply the contraction of heavy propagators
to contact interactions.}
\endinsert

At tree level, $\Phi$ exchange generates a $\phi^4$ interaction in the
effective theory, so we find that
$$\CL_{eff}^0 = \half (\partial \phi)^2 -\half m^2\phi^2 - c_0 \(\kappa\over
M\)^2 {\phi^4\over 4!} +\ldots$$
where the number $c_0$ is dimensionless and $\CO(1)$, and computable from the
graphs.  The $\ldots$ refers   to operators  such as $(\kappa^2/M^4)\phi^2
\partial^2\phi^2$ that one finds expanding the one-$\Phi$ exchange diagram to
order $p^2$.  If I had included a $\Phi^3$ interaction in the full theory,
there would have been more complicated tree diagrams leading to operators with
higher powers of $\phi$ in $\CL_{eff}$.  The tree level matching condition is
shown at the top of fig. 4.  The graphs on the left are $\Phi$ exchange graphs
in the full theory, while the contact interaction on the right is a local
operator in the effective theory. For nonrelativistic $\phi$ particles, the
procedure is equivalent to replacing the short range Yukawa potential due to
$\Phi$ exchange with a $\delta^3(\vec r)$ potential with a suitably matched
coefficient.

Now consider matching at $\CO(\hbar)$.  We must consider graphs with both 2 and
4 external $\phi$ fields.  First consider the ones with two external $\phi$
fields.  The  mass renormalization graphs are divergent in both theories and
are computed in $\bar {MS}$  To avoid large logarithms in the matching
conditions of the form $M^2\ln M^2/\mu^2$ we choose the renormalization scale
$\mu=M$.  Then the loop graphs in the second line of fig. 4 are well defined,
finite objects, and the equation defines the $\CO(\hbar)$ $\phi^2$ interactions
of the effective theory, labeled by a ``1'' on the right side of the equation.
Including these terms, the kinetic term of $\CL_{eff}$ becomes
$$\half\(1 + a_1 {\kappa^2\over 16\pi^2 M^2} \) (\partial \phi)^2 - \half\(m^2
+ b_1{\kappa^2\over 16\pi^2}\) \phi^2$$
where $a_1$ and $b_1$ are again dimensionless, $\CO(1)$, and computable from
the graphs.  I have explicitly pulled out of the graphs the dimensionful
quantities and the factors of $1/16\pi^2$ that arise from the loop integration.

Some of the graphs with four external $\phi$'s  are shown
on the third line in fig. 4.  With zero external momentum, the graphs are
approximately equal to $\kappa^4/(16\pi^2 M^4)$ times logarithms.  The
logarithms blow up in the limit that the $\phi$ mass $m$ goes to zero (an
``infrared divergence'').  However, the loop graphs in the effective theory
have
the {\it exact} same infrared divergence.  Therefore the $\CO(\hbar)$
contribution to a $\phi^4$ interaction in the effective theory (labelled by a
``1'' on the last line of fig. 4) does not blow up as $m\to 0$.  After 1-loop
matching has been performed, the effective theory looks like:
\eqn\iloop{\eqalign{
\CL_{eff}^1 &= \half\(1 + a_1 {\kappa^2\over 16\pi^2 M^2} \) (\partial \phi)^2
- \half\(m^2 + b_1{\kappa^2\over 16\pi^2}\) \phi^2\cr
&- \[c_0\(\kappa^2\over M^2\)  + c_1 \(\kappa^4\over 16\pi^2
M^4\)\]{\phi^4\over 4!} }}
where the coefficients $a$, $b$ and $c$ are   $\CO(1)$.  In addition there are
higher dimension operaotrs, such as $\phi^6$, $(\phi\partial^2\phi)^2$, etc.
This Lagrangian can be
used to compute $2\phi\to 2\phi $ scattering up to 1 loop.  One can perform an
$a_1$-dependent rescaling of the $\phi$ field to return to a conventionally
normalized kinetic term.

Let me close this section with several comments about the above example:
\item{$\bullet$}  Notice that the loop expansion is equivalent to an expansion
in $(\kappa^2/16\pi^2 M^2)$.  To the extent that this is a small number,
perturbation theory and the loop expansion makes sense.
\item{$\bullet$} We only computed relevant operators.  There are in addition
effects that are suppressed by powers of $E^2/M^2$ in an experiment with energy
$E$ (irrelevant operators).  These may be as important as a subleading
correction to a relevant operator's coefficient.
\item{$\bullet$}  We see an example of naturalness:  the matching correction to
the scalar mass is not proportional to $m^2$, so that it is ``unnatural'' for
the physical mass to be $\ll {\kappa^2\over 16\pi^2}$ --- that would require a
finely tuned conspiracy between $m^2$ and $\kappa^2$.   For $\kappa$ and $m$
both very small there is a symmetry regained in the full theory, namely the
shift symmetry $\phi\to \phi + {\rm constant}$, which explains why $\phi$ can
be naturally light in this limit.
\item{$\bullet$} The coefficients of operators in the effective field theory
are regularization scheme dependent.  Their values differ for different
schemes, but physical predictions do not (e.g, the relative cross sections for
$2\phi\to 2\phi$ at two different energies).
\item{$\bullet$} The coefficients of operators in the effective field theory
are $\mu$ dependent, where $\mu$ is the renormalization scale. (More on this
below).
\item{$\bullet$} In the matching conditions the graphs in both theories have
pieces depending nonanalytically   on light particle masses and momenta (eg,
$\ln m^2/M^2$ or $\ln p^2/M^2$)...these terms cancel on both sides of the
matching condition so that the interactions in $\CL_{eff}$ have a local
expansion in inverse powers of $1/M$.   This is an important and generic
property of effective field theories.

\lfb\hrule
\lfb
{\bf Exercise 9.} {\sl Compute  the graphs in fig. 4, using the $\bar{MS}$
scheme, and determine the coefficients $a$, $b$, and $c$ in eq. \iloop. }
\lfb
\hrule
\lfb
{\bf Exercise 10.} {\sl Draw a graph in the full theory that is not 1LPI (``one
 light particle irreducible'') and convince yourself that that it is included
in the effective theory, provided one matches all 1LPI graphs. }
\lfb
\hrule

\newsec{Quantum corrections: the myth of marginality}

We have seen that relevant interactions --- those with dimension $<4$ (or $<d$
in $d$ dimensions) --- dominate physics at low energies.  Marginal interactions
(dimension 4) would appear to be equally important at all scales.  In fact,
quantum corrections change the scaling dimension of operators from their
classical value.   This doesn't usually have a dramatic effect on relevant or
irrelevant operators, but for marginal operators it means that they become
either relevant or irrelevant.

\subsec{Renormalization group and $\phi^4$ theory}

To be concrete, consider $\phi^4$ theory with the Lagrangian
\eqn\piv{\CL=\half (\partial \phi)^2 - \half m^2\phi^2 -
{\lambda\over 4!}\phi^4\ .}
Consider the calculation of a   the 1PI Green's functions $\Gamma_n$, which are
one particle irreducible graphs that have had the external propagators
amputated.  They can be directly related to scattering amplitudes.  Ignoring
the issues of renormalization, one would expect to express these Green's
functions in terms of the external momenta, the particle mass $m$, and the
coupling constant $\lambda$:
$$\Gamma_n(p_1,...,p_n; m,\lambda)\ .$$
The dimension of this object\foot{$\Gamma_n$ is  the time ordered product of
$n$ scalar fields ($ d=n$) Fourier transformed to momentum space ($ d=-4n$)
with $n$ external propagators removed ($ d=2n$) and a factor of
$\delta^4(p_{tot})$  factored out ($ d=4$)...this gives $d= 4 -n$.} is $(4-n)$.
 Therefore if one scales all of the external momenta by a factor $s$, one
expects
\eqn\engscale{\Gamma_n(sp;m,\lambda) = s^{4-n} \Gamma_n(p;m/s,\lambda)\ .}
This expresses precisely what I was saying earlier about how the scalar mass is
a relevant operator --- note that its effects become large for small momentum
scales, corresponding to $s\ll 1$.  On the other hand, the $\phi^4$
interaction's marginality  is the observation that the importance of the
$\lambda$ coupling is independent of scale.

This analysis is incorrect when quantum corrections are taken into account, due
to the introduction of a new scale $\mu$.  When we compute in perturbation
theory, we must include counterterms and define the renormalized Lagrangian
\foot{See Ramond's book \ramond\ for details;  also see David Gross' 1975 Les
Houches lecture \ref\dgros{D. Gross, in {\it Methods of Field Theory}, ed. R.
Balian, J. Zinn-Justin, North-Holland/ World Scientific (1981)}.}
$$\CL_{ren.} = \CL + \CL_{ct}$$
where $$\eqalign{
\CL_{ren} &= \half (\partial \phi_0)^2 - \half m_0^2\phi_0^2 - {\lambda_0\over
4!}\phi_0^4€\cr
\CL&=\half (\partial \phi)^2 - \half m^2\phi^2 -
\mu^{2\epsilon}{\lambda\over 4!}\phi^4\ ,\cr
\CL_{ct} &= \half A(\partial \phi)^2 - \half m^2B\phi^2 -
\mu^{2\epsilon}{\lambda\over 4!}C\phi^4\ .€\cr}$$
Both $\CL$ and $\CL_{ct}$ must be regulated; here I have chosen dimensional
regularization, and a factor of $\mu^{2\epsilon}$ is inserted to keep $\lambda$
dimensionless, where $\mu$ is the arbitrary renormalization scale.
The Lagrangian $\CL$ is written in terms of finite parameters, but gives
infinite results;  $\CL_{ct}$ gives the counterterms $A$, $B$, $C$ which all
have $1/\epsilon$ poles in dimensional regularization and blow up in the
$\epsilon\to 0$ limit.  Computing graphs with the sum $\CL_{ren} = \CL +
\CL_{ct}$, which is written in terms of ``bare'' couplings and fields,  yields
finite answers.  The obvious correspondence between bare and renormalized
parameters is:
$$\phi_0 = \sqrt{1+A}\,\phi\equiv \sqrt{Z_\phi}\,\phi\ ,\qquad m_0^2 = m^2(1+
B)/Z_\phi\ ,\qquad \lambda_0 = \lambda(1+C)/Z_\phi^2\ .$$
We can treat $\lambda_0$, $m_0$, $\mu$ and $\epsilon$ as independent
parameters, and express $\lambda$ and $m$ in terms of them.

We can now define either bare or renormalized Green's functions, $\Gamma^0$ and
$\Gamma$ respectively.  The relation between the two is
$$\Gamma^0_n(p_1,...,p_n;\lambda_0,m_0,\epsilon) =
Z_{\phi}^{-n/2}\Gamma_n(p_1,...,p_n;\lambda_,m,\mu,\epsilon)$$
where $\Gamma_n$ is finite as $\epsilon\to 0$.   Using the fact that
$\Gamma_n^0$ is independent of $\mu$, so that ${{\rm d}\Gamma_n^0 / {\rm d}\mu}
=0$, one can derive the renormalization group (RG) equation
\eqn\rg{ \[\mu {\partial\ \over\partial\mu} + \beta  {\partial\
\over\partial\lambda} + \gamma_m m {\partial\ \over\partial m} - n\gamma\]
\Gamma_n =0\ ,}
where $\beta = \mu\partial\lambda/\partial\mu$,  $\gamma_m = \mu\partial
m/\partial\mu$,  $\gamma = \half\mu\partial \ln Z_\phi/\partial\mu$.  One can
compute these functions in perturbation theory by relating $m$, $Z_\phi$ and
$\lambda$ to $m_0$, $\lambda_0$ and $\mu$ and $\epsilon$.  For $\phi^4$ theory
one finds to leading nonzero order in perturbation theory
\eqn\rgphi{\beta(\lambda) = {3\lambda^2\over {16 \pi^2}}\ ,\qquad \gamma_m =
{\lambda\over 16\pi^2}\ ,\qquad \gamma = {1\over 12} \({\lambda\over
16\pi^2}\)^2\ .}

The reason why the RG equation is useful is because it tells one what happens
if one scales the external momenta, given that there is a new scale in the
problem, $\mu$.      On rescaling momenta by $s$, eq. \engscale\ must be
modified to read
 \eqn\sequi{\Gamma_n(sp;m,\lambda,\mu) =
s^{4-n}\Gamma_n(p;m/s,\lambda,\mu/s)}
or equivalently
\eqn\sequii{\[s {\partial\ \over\partial s} + m {\partial\ \over\partial m} +
\mu {\partial\ \over\partial\mu} - (4-n)\] \Gamma(sp;m,\lambda,\mu)=0.}
This can be combined with the renormalization group equation \rg\ to yield
an equation which relates the scaling of $s$ to changes in $m$ and $\lambda$
alone, and not $\mu$:
\eqn\scal{\[-s {\partial\ \over\partial s}+ \beta  {\partial\
\over\partial\lambda} + (\gamma_m-1) m {\partial\ \over\partial m}-n\gamma
+4-n\]\Gamma_n(sp;m,\lambda,\mu)=0.}

{\it If} one uses a mass independent subtraction scheme such as $\bar{MS}$,
then the coefficients $\beta$, $\gamma_m$ and $\gamma$ depend only on $\lambda$
and not on the other dimensionless quantity, $m/\mu$.  In this case, one can
solve eq. \scal, and one finds
\eqn\rgres{
\Gamma_n(sp;m,\lambda,\mu) = s^{4-n} \Gamma_n(p;\bar m(s),\bar\lambda(s),\mu)
e^{-n\int_1^s {\rm d}s' \gamma(\bar\lambda(s'))/s'}}
where $\bar\lambda$ and $\bar m$ satisfy the differential equations
$$ s{\partial\bar\lambda(s)\over\partial s} = \beta(\bar\lambda(s))\ ,\qquad
\bar\lambda(1)=\lambda$$
$$ s{\partial\bar m(s)\over\partial s} = (\gamma_m-1)\bar m(s)\ ,\qquad\bar
m(1)=m\
.$$

First look at this solution at tree level, where $\beta=\gamma_m=\gamma=0$ and
$\Gamma$ is independent of $\mu$.  Then the solution \rgres\ is equivalent to
the simple scaling property \sequi.  If only $\gamma$ is nonzero, and it is
constant, then the exponential in eq. \rgres\ gives an overall factor of
$s^{-n\gamma}$ to the scaling of $\Gamma$...the engineering dimension $(4-n)$
is modified by an additional factor of $-\gamma$ for each of the $n$ fields,
hence the name ``anomalous dimension'' for $\gamma$.  Finally, if $\beta$ and
$\gamma_m$ are nonzero, then changing the momentum scale means one lets the
mass and coupling ``run''.  Using the $\beta$ function in eq. \rgphi, one finds
$$\beta(\lambda) = {3\lambda^2\over {16 \pi^2}}\quad\longrightarrow\quad
\bar\lambda(s) = {\lambda\over 1- (3\lambda/16\pi^2) \ln s}\ .$$
See fig. 5.
\topinsert
\centerline{\epsfysize=1in€\epsfbox{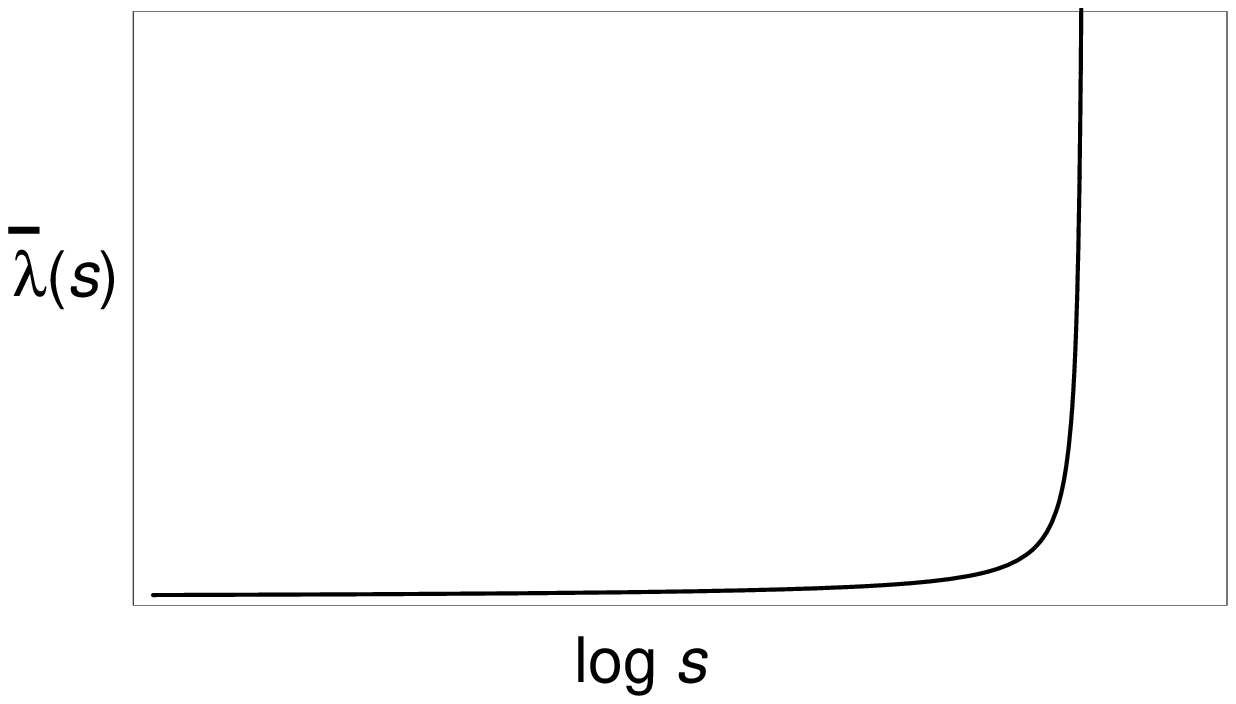}}
\caption{Fig. 5. The solution for the running coupling $\bar\lambda(s)$ as a
function of $\ln(s)$.  The one-loop expression becomes infinite at finite $\ln
(s)=16\pi^2/3\lambda^2$, but the result is not to be trusted since the
perturbative expansion breaks down.}
\endinsert

We see that the $\phi^4$ interaction is an example of a marginal interaction
that becomes irrelevant due to quantum corrections:   the lower the energy
scale
probed in a scattering experiment, the weaker the effect of the interaction.
QED is another example --- the gauge interaction becomes irrelevant due to
quantum corrections.   There is a simple physical explanation for this:  the
vacuum acts as a dielectric, with virtual particle-antiparticle pairs which
screen charges.  The greater the impact parameter in a scattering experiment,
the more screened the charge is and the weaker the interaction.  This can be
parametrized by a scale dependent fine structure constant, $\alpha(\mu)$.  As
$\mu\to 0$, $\alpha(\mu)\to 0$.  In QED, the screening ceases over distances
longer than the Compton wavelength of the electron, and so $\alpha(\mu) \to
1/137$ for $\mu\ltap m_e$.   Theories such as QED and $\phi^4$ all by
themselves are called ``asymptotically unfree''.  they are thought to be
meaningless as theories because of what happens in the ultraviolet:  In
$\phi^4$ theory one finds nonperturbatively (ie, on the lattice) that
$\lambda(\mu)\to \infty$ for $\mu\to \mu_0$ for a
finite $\mu_0$.  QED probably behaves similarly, although people debate whether
$\alpha$ may approach a constant for sufficiently large $\mu$ (a ``nontrivial
fixed-point'').

\subsec{Renormalization group and QCD}

In contrast, Yang-Mills theories such as QCD have a negative $\beta$-function
and are asymptotically free:  the gauge interactions, which are marginal at
tree level, become relevant.  The important physical difference between QED and
Yang-Mills theories that accounts for the different sign of the
$\beta$-function is  that Yang-Mills gauge bosons carry charge, while photons
do not.  For QCD, the $\beta$ function at one loop order with $N_f$ flavors of
(Dirac) quarks is
\eqn\bqcd{\beta(g) = \mu{\partial g\over \partial \mu} = {g^3\over
16\pi^2}\[-11 + {2N_f\over 3}\]\equiv -{b_0 g^3\over 2}\ .}
For $N_f\le 16$ this is negative, and so it is negative in the standard model
where $N_f=6$ (u,d,s,c,b,t).  Defining $\alpha_s = g^2/4\pi$, eq. \bqcd\ can be
integrated to
give
$$\alpha_s(\mu) = {1\over  1/\alpha_s(\mu_0)+4\pi b_0\ln(\mu/\mu_0)} \equiv
{1\over  4\pi b_0 \ln(\mu/\Lambda_{QCD})}\ .$$
Notice that a new scale has crept into the theory --- $\Lambda_{QCD}$.  It has
been defined as
\eqn\lamdef{\Lambda_{QCD} = \mu_0 E^{1/4\pi b_0\alpha(\mu_0)}\ ,}
and is independent of $\mu_0$ (to the order we are working in perturbation
theory). This is
the scale that determines where the strong interactions get strong, what the
proton and $\rho$ masses are, etc.  It's value is scheme dependent, and
depending on which sensible scheme one uses, it can range from $100 - 250$ MeV
\foot{One does not determine it by measuring where $\alpha_s$ blows up!
Instead one determines $\alpha_s$ at some large scale, such as the $Z$ mass,
where $QCD$ is weakly coupled  and a perturbative calculation of $\beta$ makes
sense. Then $\Lambda$ is defined (at one-loop order) by eq. \lamdef.}.
See fig. 6 for a plot of $\alpha_s(\mu)$.
\topinsert
\centerline{\epsfysize=1in€\epsfbox{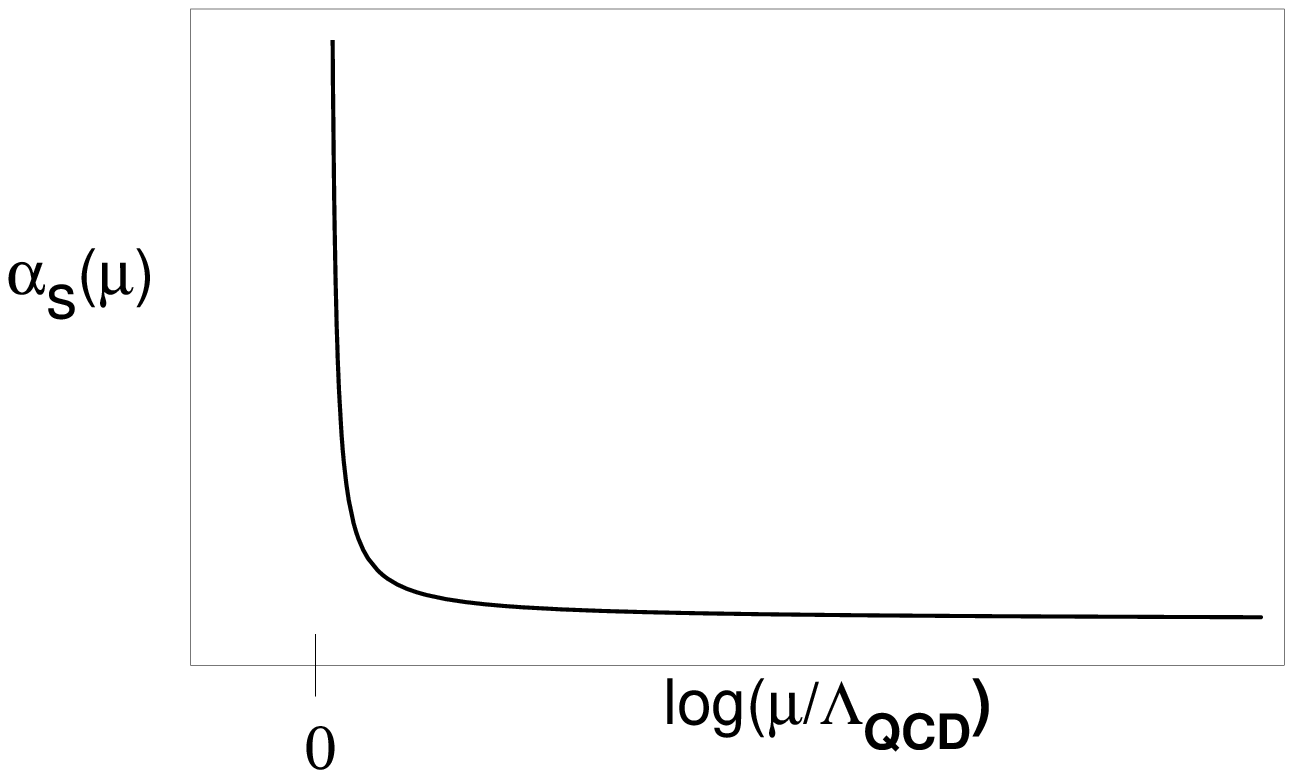}}
\caption{Fig. 6. The solution for the running coupling $\alpha_s(\mu)$ in QCD
using the one-loop $\beta$ function.  The behavior near $\mu=\Lambda_{QCD}$ is
unreliable where $\alpha_s$ is large.}
\endinsert

The reason why we call the strong interactions ``strong'' is because the beta
function is negative, and there are some light quarks.  Even though the the
gauge coupling $\alpha_s$ has positive dimension, it is small when $\alpha_s$
is small and the interaction is only barely relevant.  In contrast, quark
masses start off with a classical dimension 1.  Assuming for the moment that
QCD with explicit quark masses was the true theory (ie, ignoring the weak
interactions and the Higgs), let us look at the theory from the vantage point
of the Planck scale.  We would see a small $\alpha_s$ with sluggish logarithmic
scaling racing against tiny quark masses with linear scaling properties.  Which
one wins in the infrared?  For the top quark, the mass term wins ---
$\alpha_s(m_t)\simeq 0.1$ and the toponium ($\bar t t$) mass is determined by
$2m_t$ with only small ($\alpha_s^2m_t$) Coulomb corrections.  In contrast,
the gauge interaction wins in the race against   the $u$ and $d$ quarks,  which
have  masses of order 10 MeV.  The proton --- a $uud$ bound state --- has a
mass equal to 940 MeV, which is scarcely affected by the $u$ and $d$ quark
masses.  Its mass is attributable to the effects of the strong gauge
interaction, which is associated with a nonperturbative (scheme dependent)
mass scale $\Lambda_{QCD} \sim 200$ MeV (${\bar {MS}}$).   The strong
interactions are strong because gluon interaction is relevant and the $u$, $d$
and $s$ quarks are light \foot{It should seem peculiar to the reader that the
quark masses are scattered within a couple orders of magnitude of
$\Lambda_{QCD}$ --- this doesn't seem natural from the Planck scale
perspective.  In fact the situation is complicated by the fact that above the
weak scale, quarks don't have masses, but rather Yukawa couplings to the Higgs.
 The mystery then becomes, why does the Higgs get an expectation value within a
couple orders of magnitude of $\Lambda_{QCD}$?  Why do the Yukawa couplings
range from $10^{-5}$ (for the $u$ quark) to $1$ (for the top quark)?  There are
a range of explanations to these questions in the literature, with a range of
plausibility, but there is no evidence for any of them presently.}

\subsec{The RG and perturbation theory}

The $\beta$ function for marginal interactions can be computed without much
problem in perturbation theory; similarly one can see how relevant and
irrelevant operator dimensions are modified (anomalous dimensions).  My
favorite treatment of the subject --- called renormalization group analysis ---
is in the book by Ramond \ramond, \S 4.5.  Unfortunately, perturbation theory,
with its attendant divergences and counterterms, may be a practical
computational tool, but it tends to obscure the beautiful physics behind the
renormalization group.  Wilson thought of the effective action $S_\Lambda$ as a
theory with all modes of frequency $\omega>\Lambda$ removed.
$S_{\Lambda-\delta\Lambda}$ could be defined as
$$e^{-  S_{\Lambda-\delta\Lambda}} = \int_{\delta\Lambda} {\rm d}\phi_\Lambda
e^{-S_\Lambda}$$
where the integral is a path integral over all modes with frequency
$\Lambda-\delta\Lambda < \omega<\Lambda$.  Since the integration is over a
finite number of modes (assume the system is in a box), the integration is
finite and one needn't ever discuss counterterms, renormalization, etc. The
action can then be shown to obey a differential equation
$${\partial S_\Lambda\over\partial\Lambda} = F(S_\Lambda)$$
where $F$ is a functional of the action.  Couplings in the effective action
``flow'' as one changes the cutoff $\Lambda$.  Those with negative eigenvalues
are irrelevant, and their coupling flows to zero in the infrared (limit of
small $\Lambda$); positive eigenvalues correspond to relevant operators and
their effects become stronger in the infrared.   Wilson's  picture is in many
ways less confusing than the perturbative renormalization group discussion, but
it is not practical for analytic computations.

At any finite order in perturbation theory, one's answers {\it will} be $\mu$
dependent, and one has to pick a scale.  Changing $\mu\to\mu'$ corresponds to
changing the coupling constant $g  \to g'$.  If
$$\beta = bg^2$$
for example, then
$$g' = {g\over 1- b g t} = g\[1 + bgt + (bgt)^2+\ldots\]$$
where $t=\ln\mu/\mu'$.    So a calculation to second order in $g(\mu')$
includes an infinite number of terms in a perturbative expansion in
$g(\mu)\ln\mu/\mu'$.  Scaling $g$ is said to ``sum the leading logs''. Using
the two loop $\beta$ function sums terms like $(g^2 t)^n$, the ``subleading
logs''.  So using different values for $\mu$ {\it does} change results of a
calculation to some finite order in $g$.  In practice then one wants to choose
the value for $\mu$ that makes the perturbation expansion converge most
quickly.
Typically, that means choosing $\mu$ to make the logs small, which means
choosing $\mu$ to be a physical scale in the process of interest, $eg$ the
$q^2$ flowing through the graph.  This is good news if you are doing QCD at
$q^2=(100\ GeV)^2$, since $\alpha_s(\mu)\sim 0.1$ at that scale; it is bad news
if
you wish to do a QCD calculation at $q^2=(500\ MeV)^2$ since there is no
perturbative expansion in $\alpha_s$ at that low value for $\mu$.  To figure
out the right
value of $\mu$ one really needs to compute next order corrections and find the
$\mu$ that minimizes them.  There are interesting prescriptions for choosing
$\mu$ in the literature \ref\blm{S. Brodsky, P. LePage, P.B. MacKenzie,
\physrev{D28}{1983}{228}}.

The procedure for computing photon-photon scattering at $q^2 \ll m_e^2$  should
be clear now:
\item{$\bullet$} Match QED to an effective theory without photons, choosing
$\mu\simeq m_e$;
\item{$\bullet$} Compute the $\beta$'s and $\gamma$'s of the effective theory;
\item{$\bullet$} Change $\mu$ to the $q^2$ scale of the process of interest,
scaling the parameters of the effective theory;
\item{$\bullet$} Compute the process of interest.

\noindent
Of course, the scaling isn't very interesting in low energy QED --- the only
interactions are irrelevant, and the lowest order 4-photon vertex does not run
with $\mu$; only higher order operators do.  Where scaling effects are
important is when an interaction is pretty strong (eg, scaling effects due to
gluons in the perturbative regime), or when the scaling is over a huge energy
range (eg, the computation of $\sin^2\theta_w$ from GUT theories.  See ref.
\ref\gqw{H. Georgi, H. Quinn, S. Weinberg, \prl{33}{1974}{451}}.).  See lecture
notes by A. Manohar from the 1995 Lake Louise Winter Institute for a nice
discussion of the scaling of $\Delta S=1,2$ operators from the weak
interactions due to gluons \ref\amrev{A. V. Manohar, Lectures presented at the
Lake Louise Winter Institute, Feb 1995, UCSD/PTH 95-07}.

\lfb
\hrule
\lfb
{\bf Exercise 11.} {\sl Given that to leading nonzero order the renormalization
parameters $C$ and $Z_\phi$ in eq. \rgphi\ for $\phi^4$ theory ($\bar{MS}$) are
 $C=-\frac{3\lambda}{32\pi^2}\frac{1}{\epsilon}$ and $Z_\phi = 1 -
\frac{\lambda^2}{384\pi^2}\frac{1}{\epsilon}$, show that $\beta$ is as given in
eq. \rgphi.  Hint: use the fact that $\lambda_0$ is independent of  $\mu$, work
consistently to a given order in perturbation theory, and only set $\epsilon\to
0$ at the end of the calculation). }
\lfb
\hrule
\lfb
{\bf Exercise 12.} {\sl Compute $\alpha_s$ at $\mu$= 2 GeV using the one-loop
$\beta$ function for QCD, given that $\alpha_s$ at $\mu=90$ GeV equals $0.12$.
For the sake of this calculation, assume that the $b$ quark mass is $m_b = 4.5$
GeV.  All the other quarks are lighter than 2 GeV, except the top, which is
heavier than 90 GeV.  Assume there are no other colored particles.}
\lfb
\hrule
\lfb
{\bf Exercise 13.} {\sl Explain how you know that the four photon vertex
doesn't run with $\mu$ in effective QED below $m_e$, in the $\bar{MS}$ scheme.}
\lfb
\hrule

\newsec{Effective field theory with heavy stable particles}

Previously I mentioned that one might be interested in scattering
electrons and photons at energies much below the electron mass.  One
immediately encounters a problem in constructing the effective field theory in
terms of local operators constructed out of the electron and photon fields and
their derivatives:  an operator such as $(\bar e \Dsl^2 e)F^2/m_e^5$ is not
actually a $1/m_e^5$ effect since the time derivatives acting on an electron at
rest bring   powers of $m_e$ into the numerator.  The solution is similar to
what we have always done in nonrelativistic physics...ignore the electron rest
mass and redefine the electron field to get rid of the $exp(-imt)$ phase and
assume that the remaining frequency --- corresponding to the kinetic energy ---
is much smaller than $m$.

A free, charged, nonrelativistic particle obeys the Schr\"odinger equation:
$$i(\partial_t -ie A_0) \Psi = -{(\vec\nabla-ie\vec A)^2\over 2 m} \Psi\ .$$
in the infinite mass limit, we can ignore the kinetic energy which goes as
$1/m$ and so $\partial_t\Psi = 0$ is the equation of motion.  We can restore
relativistic covariance by defining the four velocity vector, which equals
$(1,0,0,0)$ in the rest frame of the particle, and so the equation of motion
becomes
$$v_\mu D^\mu \Psi = 0\ ,$$
and the kinetic term in the Lagrangian is
\eqn\hfree{\CL_0 =\Psi^{\dagger} v\cdot D \Psi\ .}

How does one get from the relativistic field theory to one with a kinetic term
like \hfree?  One defines the momentum of the heavy particle to be
$$p_\mu = mv_\mu + k_\mu$$
where $k\ll m$.   Then for a scalar field, one rewrites $\phi$ as $\phi =
e^{-imv\cdot x}\Psi_v$.  Then one removes the positive frequency component of
$\Psi$ which creates antiparticles, and writes the most general Lagrangian in
terms of the negative frequency component $\Psi_v^-$ and its derivatives.  The
result is a theory which involves an expansion in $k/m$, assumed to be small.
One then integrates of $v's$ to restore Lorentz covariance.

  The  procedure for fermions   is to define \ref\hmgii{H. Georgi,
\pl{240}{1990}{447}}
$$h_v = {1+\vsl\over 2} e^{i m \vsl v\cdot x} \psi\ ,$$
where $\psi$  is the heavy fermion field.  The $(1+\vsl)/ 2$ projection
operator eliminates the ``small components'' of the spinor, which are
suppressed by $1/m$.    In the large mass limit processes do not change $v$,
and so   one then constructs the effective lagrangian $\CL_v$  out of $h_v$ and
expands in powers of $\partial_\mu/m$; then one integrates $\CL_v$ over
velocities $v$.  All applications of interest have extensive symmetries that
limit the form of the higher dimension operators that one can write down in the
effective theory.

This procedure has been used extensively over the past few years to analyze
hadrons containing a heavy ($b$ or $c$) quark by constructing an effective
theory in  powers of $m_b$ and $m_c$.  Another application has been the
interaction of pions with baryons, treating the baryons as heavy fields.  See
\ref\hqet{M. Wise, hep-ph/9311212}\ for a discussion of heavy quark effective
field theory, and   \ref\jnm{E. Jenkins, A. Manohar, UCSD/PTH 91-30, talks
presented at the workshop on ``Effective Field Theories of the Standard
Model'', Dobog\'ok\H o, Hungary, August 1991.} for the heavy baryon formalism.

\newsec{Effective field theory for the strong interactions: chiral Lagrangians
}
Symmetry is the only reason we know about that can explain why mass hierarchies
occur.  The proton mass is much lighter than the Planck scale, and we can
qualitatively understand that by noting that $m_p$ arises from spontaneous
breaking of chiral symmetry.  Chiral symmetry breaking is a nonperturbative
effect which is expected to occur at a scale $\mu e^{-a/\alpha_s}$ (see eq.
\lamdef), where
$a$ is some number and $\alpha_s$ is the strong coupling at the scale
$\mu$, which might be the Planck scale.  Then the reasonable number
$a/\alpha\simeq 40$ explains the observed (enormous) hierarchy.  Attempts have
been made to similarly explain why $M_W$ is so much lighter than the Planck
scale, but no convincing theory exists.

Chiral symmetry is a symmetry that keeps fermions light.  Symmetries can also
keep bosons light, but only if they are spontaneously broken.  Then Goldstone's
theorem guarantees that there will be massless Goldstone bosons.  If what is
broken is only an approximate symmetry, then one finds ``pseudo Goldstone
bosons'' which are light but not massless.  As you have heard in Professor
Holstein's lecture, this is the explanation for why the pion is much lighter
than the rho meson, and that one can construct an effective field theory of
pseudoscalars and baryons   to describe low energy strong interactions.
Although the hierarchy here is not too large, the chiral effective theory
pioneered by Weinberg has had many successes.  And even though it is easy to
probe physics far above its range of validity, our theoretical failings mean
that we cannot analytically compute the coupling constants of the chiral
Lagrangian by matching with QCD, but must determine them phenomenologically.

I don't have time to say much about chiral Lagrangian calculations, but I do
want to make a comment about power counting.  The pions have two mass scales
associated with them: their mass $m_\pi=140$ MeV, and their decay constant
$f_\pi$.  This is variously defined; I take
$$\eqalign{\bra{0}j^{\mu\, a}_A\ket{\pi^b} &= if_\pi p^\mu \delta^{ab}\cr
 j^{\mu\,a}_A &= \bar q \gamma^\mu \gamma_5 T^a q\qquad
T^a=\frac{\sigma^a}{2}\cr
f_\pi&\simeq 93\ MeV}$$

The leading operator in the chiral Lagrangian is
$${f^2\over 4}\Tr\partial\Sigma\partial\Sigma^{\dagger}$$
where $$\Sigma = e^{2i\pi^a T^a/f}\ .$$
This term is {\it universal}; it depends on the scale $f$ and on the symmetry
breaking pattern $SU(2)\times SU(2)€\to SU(2)$.  It does not depend on any
other detail of QCD.  In fact, the system is so highly constrained by symmetry
that one gets the same effective theory from QCD, a linear $\sigma$ model, or
the NJL model!  This makes the chiral Lagrangian one of the preeminent examples
of how an effective field theory ``loses'' information about short distance
physics.  However, the pion mass term ($\Tr M_q\Sigma$) and higher dimension
operators (eg,
$\Tr(\partial\Sigma\partial\Sigma^{\dagger})^2$) are not universal, and
measuring their coefficients tells us (indirectly) about QCD dynamics.  But
what is the mass scale these higher dimension operators are being expanded in?
if the theory is an expansion in $p_\pi/f_\pi$ then it isn't of any use in the
real world.    In fact the expansion should probably be in inverse powers of
$m_\rho$ or some higher scale.  A nice power counting scheme was developed by
Weinberg and discussed in detail by Georgi and Manohar \ref\pcount{S. Weinberg,
Physica A96 (1979) 327; A. Manohar, H. Georgi, \np{234}{1984}{189}}.
They argue that the ``natural'' scale for the derivative expansion in powers of
$\partial/\Lambda$ is $\Lambda \ltap 4\pi f_\pi$.  The argument is based on the
requirement that the coefficient of an operator receive radiative corrections
no larger than the tree level value.  In the real world, it seems that
$\Lambda\simeq 4\pi f_\pi$ works pretty well, and so chiral perturbation theory
(ie, exploitation of the derivative expansion) works pretty well for pions, up
to $\sim 500$ MeV in some channels.  Certain features work well for kaons as
well.  Chiral Lagrangians have been applied to nuclear matter for both pion and
kaon condensation (eg, refs. \ref\kcon{D.B. Kaplan, A.E. Nelson,
\pl{175}{1986}{57}; H.D. Politzer, M. Wise \pl{273}{1991}{156}}) and nuclear
forces \ref\nfor{S. Weinberg, \pl{251}{1990}{288}; C. Ordonez \etal,
\pl{291}{1992}{459}}.

\newsec{Conclusions: Why effective field theory?}

Effective field theory is a useful tool to be learned.  Using effective field
theory makes computations simpler --- one needn't compute features of a
complicated field theory that rare of no interest in low energy physics; and in
conjunction with the renormalization group one can simply solve problems that
involve disparate scales. To learn how to do this one must work through
examples.  Here are a handful of effective field theory calculations in the
literature which I think are instructive:

\item{1.}  Renormalization group calculation of $\sin^2\theta_w$ from a theory
at $10^{14}$ GeV: ref. \gqw.
\item{2.} The matching and renormalization group scaling of $\Delta S=1$
operators from the weak scale down to the hadronic scale: ref. \ref\gw{F.
Gilman, M. Wise, \physrev{D20}{1979}{2392},\physrev{D27}{1983}{1128}}.
\item{3.} Computation of the charmonium binding energy in nuclei: ref. \charm.
\item{4.} Parity violating operators for nuclear physics in the chiral
Lagrangian: ref. \ref\pviol{D.B. Kaplan, M. Savage, \npa{556}{1993}{653}}.
\item{5.} Chiral perturbation theory with heavy baryons: ref. \ref\hbxpt{E.
Jenkins, A. Manohar, \pl{259}{1991}{353}}.
\item{6.}  Fitting properties of the $\Lambda(1405)$ to experiment at the
1-loop level in chiral perturbation theory: ref. \ref\msav{M. Savage,
\pl{331}{1994}{411}}.
\item{7.} Chiral perturbation theory for hadrons with a heavy quark: ref.
\ref\bgrin{B. Grinstein {\it et al.}, \np{380}{1992}{376}}.

\noindent
In addition, there are three other recent reviews I recommend on effective
field theories, all with quite different content, refs. \pol, \hmg, and \amrev.

Effective  field theory is much more than useful tool, however --- it is a
paradigm for considering all of physics, illuminating the reason why physics
looks ``simple'':  To a first approximation we needn't understand quantum
gravity to understand the top quark; we needn't know about the top quark to
understand the hydrogen atom;  the details of atomic structure are irrelevant
for hydrodynamics; and we needn't understand hydrodynamics to compute the
orbits of the celestial bodies.   Some may hungrily await the final theory of
everything, but effective field theory allows others of us to  take small
bites of something in the meanwhile.

\hfill
\centerline{{\bf Acknowledgements}}

I wish to thank M. Savage for useful conversations.  This work was supported in
part by DOE contracts DOE-ER-40561 and DE-FG06-91-ER40614, NSF Presidential
Young Investigator award \pyidk, and a grant from the Sloan Foundation.

\vfill\eject
\appendix {A}{Dimensional Regularization Formulas}
\def\dnk{{\rm d}^n k\ }
\bigskip
\bigskip
Consider the following integral in $n$ dimensions with a $Euclidian$ metric:
$$I_1 \equiv \int \dnk {1 \over (k^2 + a^2)^r} \ .$$
We may evaluate this making in terms of the $\Gamma$ function:
$$ \alpha^{-s} \Gamma(s) = \int_0^{\infty} {\rm d}x\ x^{s-1}\ {\rm
e}^{-\alpha x}.$$
Then
$$\eqalign{
I_1 &= {1\over \Gamma(r)} \int \dnk \int_0^{\infty} {\rm d}x\ x^{r-1}\
{\rm e}^{-x(k^2+a^2)} \cr
&={\pi^{n/2}\over \Gamma(r)} \int_0^{\infty} {\rm d}x\ x^{r-1-n/2}\
{\rm e}^{-x a^2} \cr
&= \pi^{n/2} a^{n-2r} {\Gamma(r-n/2) \over \Gamma(r)} }$$
\bigskip
Another useful integral is
$$I_2 \equiv \int \dnk {k^2 \over (k^2 + a^2)^r} \ .$$
To get this we define
$$\eqalign{
I_1(\alpha) &\equiv \int \dnk {1 \over (\alpha k^2 + a^2)^r} \cr
&= \alpha^{-n/2} I_1}\ ;$$
then by differentiating by $\alpha$ and setting $\alpha=1$ we find
$$I_2 = {n \pi^{n/2} a^{n-2r+2} \over 2(r-1)} {\Gamma (r-1-n/2) \over
\Gamma (r-1)}\ .$$
\bigskip
Finally note that
$$\eqalign{
I_3^{\mu \nu} &\equiv \int \dnk {k^{\mu} k^{\nu} \over (k^2 + a^2)^r} \cr
&= {\delta^{\mu \nu} \over n} I_2}\  .$$
\bigskip
\vfill
\eject
{\centerline {\bf Some Properties of $\Gamma$ Functions}}
\bigskip
$\Gamma$ functions have the property $\Gamma(z+1) = z \Gamma(z)$, with
$\Gamma(1)=1$.  Thus for integers $n \ge 1$,
$$\Gamma(n+1) = n!,\qquad n\ge 1\ .$$
Also useful is the value
$$\Gamma(\half) = \sqrt{\pi}\ .$$
\bigskip
The $\Gamma$ function is singular for non-positive integer arguments. Near
these singularities it can be expanded as
$$\Gamma(-n+\epsilon) ={(-1)^n\over n \!}\left[ {1\over\epsilon}+ \psi(n+1)
+\CO(\epsilon) \right]\ ,$$
where
$$\eqalign{
\psi(n+1) &= 1 + {1\over2} + \ldots + {1\over n} - \gamma\ ,\cr
	\gamma &= 0.5772 \ldots}$$
In particular,
$$\eqalign{
\Gamma(\epsilon-1) &= -{1\over\epsilon} +\gamma -1\cr
\Gamma(\epsilon) &= {1\over\epsilon} - \gamma }$$
\bigskip
\hrule
\bigskip
\centerline{Useful consequences:}
\medskip
\eqn\inti{\mu^{2\epsilon} \int {{\rm d}^{4-2\epsilon} q\over
(2\pi)^{4-2\epsilon}} {1\over q^2 + m^2} = {m^2\over
16\pi^2}\left[-{1\over\epsilon} + \gamma - 1 -\ln 4\pi + \ln(m^2/\mu^2)\right]}
\medskip
\eqn\intii{\mu^{2\epsilon}\int {{\rm d}^{4-2\epsilon} q\over
(2\pi)^{4-2\epsilon}} {1\over (q^2 + m^2)^2}
= {1\over 16\pi^2}\left[{1\over\epsilon} - \gamma +\ln 4\pi -
\ln(m^2/\mu^2)\right]}

\listrefs\bye